\documentclass[letterpaper,11pt]{article}
\usepackage[letterpaper, left=1in, top=1in, right=1in, bottom=1in,nohead,includefoot]{geometry}
\usepackage{color,charter,graphicx,natbib,enumitem,amsmath,amssymb,amsfonts,titlesec,placeins,bm,pdfpages, subfig, placeins,setspace,array,appendix}
\usepackage[svgnames,dvipsnames,x11names]{xcolor} 
\RequirePackage[colorlinks,citecolor=blue,urlcolor=blue]{hyperref}

\def\today{January 31, 2026} 

\def\tibm{\widetilde{\mathbf{m}}}
\def\tibM{\widetilde{\mathbf{M}}}
\def\tin{\widetilde{n}}
\def\tis{\widetilde{s}}
\def\tip{\widetilde{p}}

\def\blu{\color{RoyalBlue4}}

\def\tibm{\widetilde{\mathbf{m}}}\def\tibM{\widetilde{\mathbf{M}}}
\def\tin{\widetilde{n}}\def\tis{\widetilde{s}}\def\tip{\widetilde{p}}

\def\bzero{\mathbf{0}}

\def\A{\mathbf{A}}\def\B{\mathbf{B}}\def\D{\mathbf{D}}\def\G{\mathbf{G}}
\def\I{\mathbf{I}}\def\L{\mathbf{L}}\def\M{\mathbf{M}}\def\F{\mathbf{F}}\def\S{\mathbf{S}}
\def\V{\mathbf{V}}\def\W{\mathbf{W}}

\def\a{\mathbf{a}}\def\g{\mathbf{g}}
\def\m{\mathbf{m}}
\def\x{\mathbf{x}}\def\y{\mathbf{y}}\def\z{\mathbf{z}}  

\def\balpha{{\bm\alpha}}\def\btheta{{\bm\theta}} \def\bphi{{\bm\phi}}   
\def\bmeta{{\bm\eta}}\def\bgamma{{\bm\gamma}}\def\bomega{{\bm\omega}} 
\def\bLambda{{\bm\Lambda}}\def\bGamma{{\bm\Gamma}}\def\bPhi{{\bm\Phi}}\def\bSigma{{\bm\Sigma}}
\def\bOmega{{\bm\Omega}}\def\bTheta{{\bm\Theta}} \def\bmu{{\bm\mu}}\def\bnu{{\bm\nu}}

\def\tE{\textrm{E}} \def\tC{\textrm{C}}  \def\tG{\textrm{G}} \def\tN{\textrm{N}}   \def\tV{\textrm{V}}\def\tNG{\textrm{NG}}  

\def\cD{\mathcal{D}}\def\cP{\mathcal{P}}\def\cS{\mathcal{S}}

\def\seq#1#2{#1{:}#2}\def\j1J{j=\seq 1J}
\def\eqn#1{eqn.~(\ref{eq:#1})} \def\eqntwo#1#2{eqns.~(\ref{eq:#1},\ref{eq:#2})} 
\def\diag{\textrm{diag}}
\def\nar#1{\smallbreak{\narrower\noindent{#1}\par}\smallbreak}
\def\para#1{\medskip\noindent{\bf #1}}
	\def\parablu#1{\para{#1}}

\def\fbA{\L}    
\def\fD{\D}
\def\fbF{\S}  
 \def\fd{d}
\def\fphi{\eta}       
\def\fbphi{\bmeta} 
\def\GDP{\textrm{GDP}}
  
\newcommand{\blind}0 

\begin{document}\emergencystretch 3em  

\begin{center} 

{\blu\bf\LARGE Simultaneous Graphical Dynamic Modeling} 

\if0\blind
	{ \bigskip
		{\Large Luke Vrotsos \& Mike West}
	} \fi  

\bigskip\bigskip
\today
\bigskip\bigskip

\thispagestyle{empty}\setcounter{page}0

{\blu\bf \Large Abstract} 
\end{center} 
We review theory and methodology of the class of simultaneous graphical dynamic linear models (SGDLMs) that provide flexibility, parsimony and scalability of multivariate time series analysis.  Discussion includes core theoretical aspects and summaries of existing Bayesian methodology for forward filtering and forecasting with SGDLMs. The review is complemented by new theory  linking dynamic graphical and factor models, and extensions of the Bayesian methodology. This addresses graphical structure uncertainty via  model marginal likelihood evaluation, and analysis with missing data relevant to 
counterfactual analysis. The latter advances the ability to scale causal analysis to higher-dimensional time series. Aspects of the theory and methodology are exemplified in a global macroeconomic time series study with time-varying cross-series relationships and primary interests in potential causal effects. The example highlights the utility of SGDLMs with insights generated by the theoretical structure of these models, and benefits of fully Bayesian assessment of post-intervention outcomes in causal time series studies as in prediction more generally. 
 
\vfill
\noindent{This is a revised version of the earlier 2025 manuscript entitled  {\em\blu Dynamic graphical models: Theory, structure and counterfactual forecasting}

\bigskip

{
     \small Affiliations and contact information:
	     \\ \indent\indent{\blu Luke Vrotsos},  PhD student in Statistical Science
	     \\ \indent\indent\indent \href{mailto:luke.vrotsos@duke.edu}{luke.vrotsos@duke.edu}
		 \\ \indent\indent{\blu Mike West}, The Arts \& Sciences Distinguished Professor Emeritus of Statistics \& Decision Sciences   \\ \indent\indent\indent \href{mailto:mike.west@duke.edu}{mike.west@duke.edu}
   
					     \medskip\indent\indent Department of Statistical Science, Duke University, Durham, NC 27708-0251, U.S.A. 
	\normalsize
} 

\newpage					        
					        
\setstretch{1.1} 


\section{Introduction\label{sec:Intro}}

The framework of simultaneous graphical dynamic linear models (SGDLMs:~\citealp{GruberWest2016,GruberWest2017}) defines a broad class of flexible, scalable multivariate time series structures amenable to efficient Bayesian computations for sequential filtering and forecasting. 
The models combine customised representations of univariate series with sparse, time-varying representations of cross-series relationships. Bayesian forecasting with SGDLMs is increasingly seen in areas such as 
financial portfolio analysis~\citep[e.g.][]{GriveauEtAl2020, Kyakutwika2023} with related developments of efficient software~\citep[e.g.][]{GruberSGDLMrCode,pySGDLM,SAS-SGDLMs}.

For a $q-$vector time series $\y_t,$ a model in the SGDLM class has the form
\begin{equation} \label{eq:SGDLM-y-mvn}
(\I -\bGamma_t )\y_t = \bmu_t + \bnu_t\ \ \textrm{with}\ \ \bnu_t \sim \tN(\bzero, \bLambda_t^{-1})
\end{equation} 
where:   
(i) the $q-$vector $\bmu_t$ has terms such as trends and regressions on external predictors; 
(ii) the $q\times q$ matrix $\bGamma_t$ is sparse and has zero diagonal elements; 
(iii)  $\bLambda_t$ is $q\times q$ diagonal with positive residual volatilities as diagonal elements.   These structural models include time-varying vector autoregressions and related multivariate volatility models as special cases.  Such forms have been widely used in Bayesian macroeconomics~\citep[e.g.][]{Primiceri05,BanburaEtAl2010,KoopKorobilis2010,KoopKorobilis2013,NakajimaWest2013JBES,ZhaoXieWest2016ASMBI,Lopesetal2022} and other areas of signal processing~\citep[e.g.][]{NakajimaWest2015DSP,NakajimaWest2017BJPS} but with $\bGamma_t$ strictly triangular. 
SGDLMs remove that constraint,  obviating the issue of dependence on a chosen order of the univariate series in $\y_t,$ and yielding more flexible and parsimonious models.  A further advance is in efficient forward simulation and variational Bayes methodology for filtering and forecasting analyses, eliminating the need for intensive and repeat batch MCMC analyses used in most of the prior literature.  

This paper presents:  (a) a review  of the theoretical structure and current Bayesian methodology for fitting, exploring and evaluating SGDLMs; (b)  some new graphical model theory defined by SGDLMs linked to factor structure in multivariate time series;  (c)  some extensions of existing methodology for fitting and using SGDLMs addressing questions of model uncertainty and missing data;  and (d) discussion of the impact of these developments for counterfactual forecasting relevant to causal analysis in time series.   Section 2 discusses theoretical features of SGDLMs and highlights practical relevance related to underlying, graph-theoretical structure. Section 3 introduces and explores new theory linking graphical and factor models.  Section 4 reviews and summarizes existing methodology of Bayesian filtering and forecasting analyses in SGDLMs, highlighting the relevance of the new theory of Section 2 and adding new developments related to marginal likelihood evaluation for model and hyper-parameter comparisons.  Section 5 discusses causal forecasting and develops key extensions of the SGDLM methodology for the setting of partial observations central to counterfactual analyses.   Section 6  explores and illustrates the structure and theory underlying SGDLMs in detailed applied example. Section 7 builds on this applied setting to explore and illustrate the use of SGDLMs in counterfactual forecasting for observational time series studies with putatively causal interests. Section 8 adds summary comments. Appendices provide supporting theoretical and technical details.   

\section{Graphical Model Structure\label{sec:Theory}} 
 
The theoretical structure is discussed in this section without the time $t$ index, for clarity.   
For a single observation $\y=(y_1,\ldots,y_q)'$, \eqn{SGDLM-y-mvn} has the structural form  $\y=\bmu + \bGamma\y+\bnu$ with 
$\tV(\bnu)=\V=\bLambda^{-1}.$ Assume all parameters are known.  Given a  pattern of zeros in $\bGamma,$ each  $y_i$ is regressed on only those other $y_j$ for which the $i{,}j$ element $\gamma_{ij}$ of $\bGamma$ is non-zero. This is reflected in the directed (but usually {\em not} acyclic) graph where this set of the $y_j$ are {\em simultaneous parents} of $y_i$, denoted by $sp(i)=\{ j=\seq 1q, j\ne i: \ \gamma_{ij}\ne 0\}.$   This $y_i$ is a {\em child} of each of its parents; the {\em child set} of any $y_j$ is $ch(j) =  \{ i=\seq 1q, i\ne j: \ \gamma_{ij}\ne 0\}.$  Then, $\I-\bGamma$ is non-singular~(see Appendix~\ref{app:eigenGamma} that exploits aspects of graph theory). Hence 
$\y \sim \tN(\balpha,\bOmega^{-1})$ with mean $\balpha=(\I-\bGamma)^{-1}\bmu$
and precision matrix $\bOmega=(\I-\bGamma')\bLambda(\I-\bGamma).$

\subsection{Impact of Parental Specification\label{sec:ImpactofParents}}

\noindent{\bf Cross-Talk in Mean Predictions.} The point prediction $\alpha_i$ of $y_i$ in  $\balpha=(\I-\bGamma)^{-1}\bmu$ will generally have terms involving $\mu_j$ for $j\ne i$ as well as $\mu_i,$ i.e.,   
\lq\lq spillover''  from some other series. Insights into this are easily appreciated in cases (typical in practice-- see Appendix~\ref{app:eigenGamma}) that $\max(|g_i|)<1$
where $g_{\seq 1q}$ are the eigenvalues of $\bGamma.$ We then have the convergent infinite series representation
$$ \alpha_i = \mu_i + \sum_{j\in sp(i)} \gamma_{ij}\mu_j 
                      + \sum_{\substack{j\in sp(i) \\ k\in sp(j)}} \{ \gamma_{ij}\gamma_{jk} \} \mu_k 
                       + \sum_{\substack{j\in sp(i)\\ k\in sp(j)\\ h\in sp(k)}} \{ \gamma_{ij}\gamma_{jk}\gamma_{kh} \} \mu_h 
                       + \cdots.
$$
This shows the cascade of effects through parental generations.  Direct parents have main effects: $\mu_j$ contributes to $\alpha_i$ with coefficient $\gamma_{ij}$ when $j\in sp(i).$  Parents of parents, i.e., grandparents, have second-order effects: $\mu_k$ contributes to $\alpha_i$ with a coefficient that sums the products 
$\gamma_{ij}\gamma_{jk}$ over $j\in sp(i)$ when $k\in sp(j).$    Here series $j$  is a mediating variable between $i$ and $k$ if $k\ne sp(i).$ 
A grandparent $y_k$ of $y_i$ can also be a parent; then the grandparental term is a second-order interaction relative to the main effect of series $k.$ 
Higher ancestors in the infinite sum have inevitably smaller contributions; successive terms have products of 3, then 4, etc. of the $\gamma_{**}$ coefficients, which decay to zero at higher generations (since the infinite sum converges).    

\parablu{Simultaneous Conditional Dependencies.} The precision matrix $\bOmega$ will typically be sparse, increasingly so with higher sparsity in $\bGamma.$ The existence and pattern of off-diagonal zeroes  in $\bOmega$ relates to the parental structure in the implied, unique undirected graphical model for $p(\y).$ A zero off-diagonal element $\omega_{ij}$ implies complete conditional independence of $y_i,y_j$ given all other series.  We have
 $\bOmega = \bLambda-\bGamma'\bLambda-\bLambda\bGamma + \bGamma'\bLambda\bGamma.$ 
The terms $\bGamma'\bLambda$ and $\bLambda\bGamma$ contribute to a non-zero $\omega_{ij}$ if either $i\in sp(j)$ or $j\in sp(i),$ or both. The final term $\bGamma'\bLambda\bGamma$ contributes a non-zero term if $y_i,y_j$ are common parents of any other series $y_k$, i.e., acts to introduce moralizing edges (marrying any unmarried pairs of parents). 
 
The off-diagonal zero/non-zero pattern in $\bOmega$  exhibits dependencies in the  {\em complete conditional p.d.f.s} $p(y_i|\y_{-i}) \propto p(\y)$   implied by the set of {\em simultaneous p.d.f.s} $p_i(y_i|\y_{sp(i)})$. For any $y_i$ that is not a simultaneous parent at all, the two are the same. For any parental series $y_i$, the complete conditional depends on its own parental and child series. Additional complete conditional dependence can arise through the moralization induced by common parents of any child as noted above.   
 
\subsection{Graph Set Intersections and Eigenstructure\label{sec:eigenstructure}}

The set of univariate series $y_i$ for which $ch(i)$ is non-empty can be partitioned into a collection of $k\le q$ {\em common parental sets} $\cP_h, h=1:k$, such that: 
(i) each series $i$ belongs to one and only one of the $\cP_h;$  
(ii) each series $i \in \cP_h$ shares at least one child with at least one of the other series 
$j \in \cP_h;$ and (iii) if series $i\in \cP_h$ then 
$ch(i)\cap ch(j)=\emptyset,$ the empty, for all series $j\ne \cP_h.$
Some of the $\cP_h$ may include a single series, while those series that are not parental predictors at all are not members of any of the $\cP_h.$ 
Any two series in the same  common parental set may not be connected as a parent/child pair themselves. 
Extend the $ch(\cdot)$ notation to subsets of series with $ch(\cP_h)$ being the set of child series of all parental series in $\cP_h.$    
Let $p_h=|\cP_h|$ where $|\cdot|$ denotes the number of series in a subset, and set $p = \sum_{h=\seq1k} p_h.$ Also, let $r_h =\min(|\cP_h|,|ch(\cP_h)|)$
 and $r=\sum_{h=\seq1k} r_h$. 
 
In the spectral decomposition $\bGamma'\bGamma=\fbF'\fD^2\fbF$, the rows of $\fbF$ are the eigenvectors of $\bGamma'\bGamma$;  the eigenvalues defining the diagonal matrix $\fD^2$ are the squares of the singular values of $\bGamma$, some of which may be zero. 
The $i{,}j$ off-diagonal element of $\bGamma'\bGamma$ is non-zero when variables $i{,}j$ are common parents of one or more children, i.e., if $ch(i)\cap ch(j) \neq \emptyset.$   With no loss of generality, order variables
so that the first $p_1$ are those in $\cP_1,$ followed by those in $\cP_2,$ and so forth, continuing to finally include any remaining series that have no child series in arbitrary order. With this ordering, $\bGamma'\bGamma = \textrm{block diag}(\bPhi_1,\ldots,\bPhi_k,\bzero)$ with 
symmetric $p_h\times p_h$ matrices $\bPhi_h$ for $h=1:k,$ followed by a $(q-p)\times(q-p)$  $\bzero$ matrix.  Here $\bPhi_h = \bGamma_h'\bGamma_h$ where $\bGamma_h$ is the matrix of all rows and columns corresponding to only $\cP_h$. Since $\bGamma_h$ has only $|ch(\cP_h)|$ non-zero rows, then $\text{rank}({\bPhi_h}) \leq |ch(\cP_h)|$; and then, as $\text{rank}({\bPhi_h}) \leq p_h$, we have
$\text{rank}({\bPhi_h}) \leq r_h$.   Each
$\bPhi_h$ may or may not have some off-diagonal zeros; the non-zeros correspond to children shared by two parents in $\cP_h.$    It is possible that $\text{rank}(\bGamma) < r$;~\citet[][Theorem 2]{fang_low-rank_2024} shows how to identify such reduced-rank cases from the graphical structure.

The rows of $\fbF$ are sparse,  each with non-zeros in columns 
corresponding to only one of the sets $\cP_h$ and with values coming from the (at most $r_h$) eigenvectors of the implied $\bPhi_h$~(see, for example, the theory in~\citealp{Simovici2014}). 
 
\section{Sparse Factor Structure\label{sec:factorstructure}}  

\subsection{Theoretic Structure\label{sec:notfactorstructure}}  

The above review is complemented here with the recognition that SGDLMs implicitly define underlying factor models. This novel 
development offers potential for common factor-related interpretations of fitted models in any application.  The structure is uncovered  by exploiting 
the singular value decomposition (SVD) of the simultaneous coefficient matrix.  

Continuing in this section without the time $t$ index, for clarity, the SVD is 
 $\bGamma = \fbA\fD\fbF$ where, with dimension $p\le q$ equal to the number of non-zero singular values,  the $q\times p$ {\em factor loadings} matrix $\fbA$ has orthonormal columns,  the $p\times q$ {\em factor scores} matrix $\fbF$ has  orthonormal rows,  
and the matrix $\fD=\diag(\fd_1,\ldots,\fd_p)'$ has the non-zero singular values of $\bGamma$. Both $\fbF$ and $\fD$ underlie the eigenstructure of $\bGamma'\bGamma$ discussed in Section~\ref{sec:eigenstructure}.  When $\bGamma$ is sparse, both $\fbA$ and $\fbF$ will generally be sparse. 

The model $\y=\bmu+\bGamma\y + \bnu$ then has the equivalent form
$ \y = \bmu +  \fbA\fbphi + \bnu $ with $\fbphi = (\fphi_1,\ldots,\fphi_p)' = \fD\fbF\y.$    This is the form of a mean-adjusted factor model with $p-$dimensional {\em factor} vector $\fbphi$ based on the  factor loadings matrix $\fbA$ and the factor scores 
matrix $\fbF$.  Each factor $\fphi_j$ is a \lq\lq linear scalar component'' defined by a subset of the variables. 
Inherently, the implied factors are dependent; $\tV(\fbphi) =\fD\fbF\bOmega^{-1}\fbF'\fD$  
may have important cross-factor correlations. This contrasts with traditional factor models that impose orthogonal factors for mathematical identification reasons that are moot here. The factors are also correlated with the residuals $\bnu$, viz. 
$\tC(\fbphi,\bnu) = \fD\fbF\A\V$  where $\A=(\I-\bGamma)^{-1}$ and $\V$ is the diagonal matrix 
$\V=\bLambda^{-1}.$   While $\tC(\fbphi,\bnu)$ will 
typically be very sparse, it may indicate non-negligible correlations.

The theory in Section~\ref{sec:eigenstructure}  implies that number of factors is $r=\sum_{h=\seq1k} r_h$ in full-rank cases. In such cases, for each $h = 1:k$ there are $r_h$ rows of $\fbF$ with non-zero weightings only on the elements of $\cP_h$. The corresponding $r_h$ columns of $\fbA$ have non-zero entries only for the variables in $ch(\cP_h)$. In reduced-rank cases, there will be fewer than $r$ factors. 

\subsection{Sparse Dynamic Factor Structure\label{sec:dynamicfactors}} 
  
Returning to SGDLMs with time index explicit, the theory of Section~\ref{sec:notfactorstructure}  yields an implied, dynamic sparse factor model form 
$\y_t=\balpha_t + \fbA_t\fbphi_t + \bnu_t$ where, based on the SVD $\bGamma_t = \fbA_t\fD_t\fbF_t$, the factor vector is 
$\fbphi_t = \fD_t\fbF_t\y_t$. The theory of Section~\ref{sec:factorstructure} applies directly here to inform on the number of factors $p$. 
In this time series setting,  there are evident connections with \lq\lq linear scalar components'' that aim to \lq\lq reveal possibly hidden simplifying structures of the process''~\citep[e.g.][]{TiaoTsay1989}. The broader connections are with sparse and dynamic factor modeling~\citep[e.g.][and references therein]{LopesCarvalho07,NakajimaWest2013JFE,Kaufmann2019,LavineCronWest2020factorDGLMs,BolfarineEtAl2024}  
In the new perspective here, however, the factor structure is derivative of the general graphical modeling approach, and the sparsity patterns in the (time-varying) 
$q\times p$ loadings matrix $\fbA_t$  and  $p\times q$ scores matrix $\fbF_t$ are determined wholly by the choice of parental sets. 

Again from Section~\ref{sec:factorstructure},  the variance matrix $\tV(\fbphi_t) =\fD_t\fbF_t\bOmega_t^{-1}\fbF_t'\fD_t$  
may have strong cross-factor dependencies.  It is important to note that other approaches to dynamic {\em  dependent} factor modeling have shown improvements in model fit and forecasting accuracy-- and to a degree interpretations--  with models that allow cross-factor correlations; in some such cases, the data analysis infers strong factor-factor dependencies~\citep[e.g.][]{NakajimaWest2013JFE,ZhouNakajimaWest2014IJF,BeyelerKaufmann2021}.

\section{Bayesian Analysis for Fitting SGDLMs\label{sec:SGDLM}}  
 
\subsection{SGDLM Specification: Reprise\label{sec:sgdlmdefinition}}

With $\y_t=(y_{1t},\ldots,y_{qt})'$, each univariate series in~\eqn{SGDLM-y-mvn} follows a dynamic linear model (DLM) 
\begin{equation}  
y_{jt} = \F_{jt}' \btheta_{jt} + \nu_{jt} \ \  \textrm{and} \  \ 
\btheta_{jt} = \G_{jt} \btheta_{j,t-1} + \bomega_{jt}
\label{eq:SGDLMj}
\end{equation}
where the observation noise terms $\nu_{jt}$ and evolution innovations $\bomega_{jt}$ are zero mean, mutually independent, and independent across series $j$  over time $t$ and $\nu_{jt}\sim N(0,1/\lambda_{jt}).$ This is a traditional DLM with regression vector  $\F_{jt} = (\x_{jt}, \y_{sp(j),t})'$ where subvector 
$\x_{jt}$ involves predictors (constants, exogenous variables, lagged $y_\ast$ values, etc.) specific to model $j,$  while $\y_{sp(j),t}$ is a subvector of $\y_t$ with indices in the  simultaneous parental set $sp(j)$ that induce cross-series, time-varying conditional dependencies. 
The Markov-evolving state vector has partitioned form $\btheta_{jt}= (\bphi_{jt}, \bgamma_{jt})'$ conformable with that of $\F_{jt}.$  
Thus  $\F_{jt}' \btheta_{jt} = \mu_{jt} + \y_{sp(j),t}' \bgamma_{jt}$ where $\mu_{jt} =\x_{jt}'\bphi_{jt}$ is element $j$ of $\bmu_t$  
in~\eqn{SGDLM-y-mvn}.  The $(j{,}h)$ element of the $q\times q$ matrix $\bGamma_t$ has the relevant element from 
$\bgamma_{jt}$ for each $h\in sp(j)$, being zero otherwise.  Further, $\bnu_t=(\nu_{1t},\ldots,\nu_{qt})'$ with precision matrix  $\bLambda_t = \diag(\lambda_{1t}, \ldots, \lambda_{qt})$.
The rest of the model structure is from standard DLM methodology: the state evolution matrix $\G_{jt}$ is known at time $t$,  the  $\lambda_{jt}$ evolve according to standard beta-gamma Markov processes, and $\tV(\bomega_{jt})$ is defined by discount factors~\citep[e.g.][chap.~4]{GruberWest2016,PradoFerreiraWest2021}. 
 
From Section~\ref{sec:Intro}, the implied distribution for $\y_t$ is 
$\y_t \sim \tN(\balpha_t, \bOmega_t^{-1})$ with $\balpha_t = (\I - \bGamma_t)^{-1}\bmu_t$ and precision matrix $\bOmega_t  = (\I - \bGamma_t)'\bLambda_t(\I - \bGamma_t)$.  Different external predictors $\x_{jt}$ over $j$ allow customisation of individual models, while small 
simultaneous parental sets $sp(j)$ induce sparse $\bGamma_t$ and parsimonious   models via the implied sparsity of $\bOmega_t$.

\subsection{Review of Filtering and Forecasting\label{sec:summarySGDLManalysis}}  

The sequential Bayesian analysis of SGDLMs integrates standard theory and methodology of univariate DLM analysis into the multivariate simultaneous system with a decouple/recouple strategy~\citep{West2020Akaike}.  Analysis is sequential over time, with time $t-1$ to $t$ forecast-update-evolution steps that are recursively applied as time $t$ evolves. This 
combines decoupled, conditionally conjugate priors and posteriors for states and volatilities $\{\btheta_{jt},\lambda_{jt}\}$ in each time period,   with analytic updating in the corresponding univariate DLMs that are then recoupled to define full inferences on the set of dynamic parameters  $\bTheta_t= [ \btheta_{1t},\ldots,\btheta_{qt}]$ and  $\bLambda_t = \diag(\lambda_{1t}, \ldots, \lambda_{qt})$ in the multivariate system of \eqn{SGDLMj}.  Full inference and multi-step ahead forecasting at each time point exploits simple
importance sampling (IS) for prior-posterior updates and direct simulation for forecasting.  Moving posterior distributions through the state and volatility evolutions of the model at each time (i) decouples the series into their univariate models, and (ii) exploits accurate Variational Bayes (VB) steps that link importance sample posterior representations to conditionally conjugate forms that enable analytic updating. 

Summary details at each time $t$ are noted. In the traditional notation, $\cD_t$ denotes all information-- including past data up to $\y_t$-- available at time $t$.  

\subsubsection{Decoupled prior for updating at time $t$.}
 Independently across series and conditional on $\cD_{t-1},$  the inferences on the series $j$ state vector and precision in each univariate model of \eqn{SGDLMj}
 for use in updating based on time $t$ observations are constrained to be of conjugate normal-gamma form
\begin{equation}
\textrm{p.d.f.}\ \  p_{j,t-1}(\btheta_{jt},\lambda_{jt}|\cD_{t-1}): \qquad (\btheta_{jt},\lambda_{jt}|\cD_{t-1}) \sim \tNG(\m_{jt}^*,\M_{jt}^*,n_{jt}^*,s_{jt}^*), \quad j=1:q, \label{eq:timetpriorj}
\end{equation}
with all parameters superscripted $*$ known at time $t-1.$ 
The $\tNG$ notation summarises $\btheta_{jt}|\lambda_{jt} \sim \tN(\m_{jt}^*,\M_{jt}^*/(s_{jt}^*\lambda_{jt}))$ and $\lambda_{jt} \sim \tG(n_{jt}^*/2,n_{jt}^*s_{jt}^*/2)$ with 
$n_{jt}^*$ degrees of freedom and  mean $1/s_{jt}^*$. 
The  $\btheta_{jt}$ margin is multivariate T with $n_{jt}^*$ degrees of freedom, mode $\m_{jt}^*$ and scale matrix $\M_{jt}^*$. 
$ (\btheta_{jt},\lambda_{jt}|\cD_{t-1})$.
With $\bTheta_t= [ \btheta_{1t},\ldots,\btheta_{qt}]$ and  $\bLambda_t = \diag(\lambda_{1t}, \ldots, \lambda_{qt})$,  the joint prior over series  is
\begin{equation}
p_t(\bTheta_{t}, \bLambda_{t} | \cD_{t-1}) = \prod_{j=1{:}q} p_{jt}(\btheta_{jt}, \lambda_{jt}| \cD_{t-1}). \label{eq:timetfullprior}
\end{equation}
 
\subsubsection{Update to posterior and model recoupling at time $t$.}
In each of the decoupled models separately, standard conjugate posteriors are  usual updates of \eqn{timetpriorj}, namely
\begin{equation}
\textrm{p.d.f.}\ \  \tip_{jt}(\btheta_{jt},\lambda_{jt}|\cD_t): \qquad (\btheta_{jt},\lambda_{jt}|\cD_{t}) \sim \tNG(\tibm_{jt},\tibM_{jt},\tin_{jt},\tis_{jt}), \quad j=1:q, 
\label{eq:timetnaiveposterior}
\end{equation}
with defining parameters based on the usual DLM updating equations. These are 
$\tibm_{jt} = \m_{jt}^* + \a_{jt} e_{jt}$, $\tibM_{jt} = (\M_{jt}^* - \a_{jt} \a_{jt}' q_{jt}) z_{jt}$,  $ \tin_{jt} = n_{jt}^* + 1$ and $\tis_{jt} = z_{jt} s_{jt}^*$
based on the point forecast error  $e_{jt} = y_{jt} - \F_{jt}' \m_{jt}^*$, 
the adaptive coefficient vector $\a_{jt} = \M_{jt}^* \F_{jt} /q_{jt}$ where the $1-$step forecast variance factor $q_t$ is defined above,  
and volatility update factor $z_{jt} = (n_{jt}^* + e_{jt}^2/q_{jt})/(n_{jt}^* + 1)$.   

The exact joint posterior is 
\begin{equation}
p_t(\bTheta_t,\bLambda_t | \cD_t)  \ \propto  \ |\I - \bGamma_t|_{+} \  \prod_{j=1{:}q} \tip_{jt}(\btheta_{jt},\lambda_{jt}|\cD_t)
  \qquad  \propto \ |\I - \bGamma_t|_{+}\   \tip_t(\bTheta_t,\bLambda_t | \cD_t)
   \label{eq:timetexactposterior}
\end{equation}
where $\tip_t(\bTheta_t,\bLambda_t | \cD_t)=\prod_{j=1{:}q} \tip_{jt}(\btheta_{jt},\lambda_{jt}|\cD_t)$, referred to as the naive posterior, defines what is typically an excellent  approximation to $p_t(\bTheta_t,\bLambda_t | \cD_t) $.

The absolute value of the determinant of $\I - \bGamma_t$  {\em recouples} the naive conjugate posteriors, accounting 
for cross-series dependencies. Independent Monte Carlo samples from each of the $\tip_{jt}(\cdot|\cdot)$ are importance-weighted by the determinant term, defining importance sampling (IS) of the exact posterior $p(\bTheta_t,\bLambda_t | \cD_t).$   The posterior is  summarised by the IS sample and weights 
$\cS_t = \{ \bTheta_t^r,\bLambda_t^r, w_t^r\}_{r=\seq1R}$ where $R$ is the number of Monte Carlo replicates, and each sampled pair $\bTheta_t^r,\bLambda_t^r$  has importance weight $w_t^r$ (summing to 1 over $r=1:R).$

\subsubsection{Forecasting ahead at time $t$.}
Direct simulation enables forecasting from the full predictive distribution over one or more  steps ahead. For replicates $r=1:R,$  (a) sample one parameter set $\bTheta_t^r,\bLambda_t^r$ from $\cS_t$ according to the IS weights; (b) conditional on this draw, sample a corresponding pair $\bTheta_{t+1}^r,\bLambda_{t+1}^r$ from the evolution equations of the SGDLM, and recurse into the future to 
$k>1$ steps ahead; (c) at each $h=1:k,$ compute the implied normal parameters   
$(\balpha_{t+h}^r, \bOmega_{t+h}^r)$ as defined in Section~\ref{sec:sgdlmdefinition}, then draw $\y_{t+h}^r$ from that conditional 
normal distribution.   The result is a draw from predictive $p(\y_{t+1},\ldots,\y_{t+k}|\cD_t)$.   

Note that the predictive distributions sampled here are inherently mixtures of normals. That is, conditional on state parameters and volatilities, predictives are conditionally normal; averaging over the Monte Carlo samples of states and volatilities defines the representations as mixtures of (large numbers of) normals. This is useful and relevant in later discussions of addressing SGDLM analyses with missing data.

\subsubsection{Decoupling for evolution to time $t+1$.} The  posterior is decoupled into a product of conjugate forms via a  mean-field, reverse 
 variational Bayes (VB) approach. This emulates the exact posterior by the product  
$ 
q_t(\bTheta_t,\bLambda_t | \cD_t)  =\prod_{j=1{:}q} q_{jt}(\btheta_{jt},\lambda_{jt}|\cD_t)
$ 
with components
\begin{equation}
\textrm{p.d.f.}\ \  q_{jt}(\btheta_{jt},\lambda_{jt}|\cD_t): \quad (\btheta_{jt},\lambda_{jt}|\cD_t) \sim \tNG(\m_{jt},\M_{jt},n_{jt},s_{jt}), \quad j=1:q. \label{eq:timetVBposterior}
\end{equation}
The parameters of each NG component are evaluated to minimise the K\"ullback-Leibler  divergence of the product form $q(\cdot|\cdot)$ from  $p(\cdot | \cdot)$, and
are trivially calculated. 
 
\parablu{Evolution to time $t+1$.}   The evolution of the $\btheta_{jt},\lambda_{jt}$ to  $\btheta_{j,t+1},\lambda_{j,t+1}$, independently in each univariate DLM, 
follows standard univariate DLM theory. This involves a volatility discount factor $\beta\in (0,1]$ and the state innovation variance matrix $\W_{j,t+1}$ for the model \eqn{SGDLMj} at time $t+1.$ 
The latter is specified using a block discount strategy~\citep[][chap.~6]{WestHarrison1997}; $\W_{j,t+1}$ is block diagonal with 2 blocks corresponding to the partition of  $\btheta_{jt}$ in the model \eqn{SGDLMj}, i.e., $\btheta_{jt}= (\bphi_{jt}, \bgamma_{jt})'$. The blocks of $\W_{j,t+1}$  are variance matrices of the evolution innovations on the dynamic regression parameters (i) $\bphi_{jt}$, of any trends and exogenous predictors, and (ii)  $\bgamma_{jt}$, of the parental predictors. Discount factors $\delta_\phi, \delta_\gamma$ define the variance matrix blocks, and allow differing degrees of discounting on the 2 subvectors.  Note that the GDP example in the paper adopts the specification that  $\delta_\phi=\delta_\gamma=\delta.$ 
The resulting prior for time $t+1$ maintains the conjugate form of \eqntwo{timetpriorj}{timetfullprior}, with parameters evolved to 
$\m_{j,t+1}^* = \G_{j,t+1} \m_{jt}$, $\M_{j,t+1}^* = \G_{j,t+1} \M_{jt}^* \G_{j,t+1}' + \W_{j,t+1}$,  $n_{j,t+1}^* = \beta_j n_{jt}$ and $s_{j,t+1}^*=s_{jt}.$ 
This completes the update-forecast-evolve steps, and analysis moves to time $t+1.$

\subsection{Marginal Likelihoods in SGDLMs\label{sec:marglikcalc}}

A novel extension of the existing methodology addresses evaluation of model marginal likelihoods in SGDLMs; this is most
relevant in addressing model uncertainty. Comparisons of interest may be between different parental structures, exogenous predictors in univariate models, or discount factors. Marginal likelihood evaluations are based on the $1-$step predictive p.d.f.s $p(\y_t|\cD_{t-1})=\tE[p(\y_{t}|\bTheta_t,\bLambda_t,\cD_{t-1})]$ where the expectation is over the time $t$ prior $p(\bTheta_t,\bLambda_t|\cD_{t-1})$.  The marginal likelihood from data over a series of times is the product of the per time $t$ values. 

A direct approximation of $p(\y_{t}|\cD_{t-1})$ is a sample average of $p(\y_{t}|\bTheta_t,\bLambda_t,\cD_{t-1})$ over draws from the prior $p(\bTheta_t,\bLambda_t|\cD_{t-1}).$  However, a more accurate Monte Carlo approximation is based on sampling the {\em posterior} rather than the prior.  Theoretical details in Appendix~\ref{app:marglikdetails} underlie the main result that 
\begin{equation} \label{eq:marglikintegral}
p(\y_{t}|\cD_{t-1}) =  g_t(\y_{t}) 
  \prod_{j=1}^{q} p_j(y_{jt}|\y_{sp(j)t},\cD_{t-1})
\end{equation}
where
\begin{equation*}  
   g_t(\y_{t}) = 
\int |\I-\bGamma_t|_+ \prod_{j=1}^{q} p_j(\bgamma_{jt}|y_{jt},\y_{sp(j)t},\cD_{t-1}) d\bGamma_t.
\end{equation*}
Here each p.d.f.  $p_j(\bgamma_{jt}|y_{jt},\y_{sp(j)t},\cD_{t-1}) $ is  that of the marginal {\em posterior} T distribution for the subvector $\bgamma_{jt}$ in the DLM for series $j$ alone. 
This explicitly shows how the computational challenge arises only through the simultaneous parental parameters $\bGamma_t$ with the implied need to evaluate  $g_t(\y_{t}).$  The impact of this term in modifying the product of univariate p.d.f.s in \eqn{marglikintegral} will be smaller as a function of sparsity of $\bGamma_t$ (with $g_t(\y_t)\approx 1$ in very sparse cases).   Operationally, the term $g_t(\y_{t})$ is approximated by Monte Carlo integration  with a sample average of $|\I-\bGamma_t|_+$
over independent draws of the $\bgamma_{jt}$ from their decoupled marginal T posteriors.  

\section{SGDLMs in Counterfactual Forecasting \label{sec:causaltime}}

\subsection{Setting and Background: Causal Forecasting\label{sec:causalsetting}}

Causal analysis in time series is an area of increasing interest, with a main focus on counterfactual forecasting. The idea of using chosen or constructed \lq\lq synthetic'' control series  to condition counterfactual forecasts of putative \lq\lq treated'' or \lq\lq experimental'' series~\citep[from][]{Abadie2003} is increasingly widespread in economics and other social sciences. Bayesian versions were introduced by \cite{Brodersen2015} for a univariate outcome series, with later work addressing steps into multivariate time series ~\citep[e.g.][]{MenchettiBojinov2022, AntonelliBeck2023}.    To date, there has been limited development of multivariate time series in such settings. This section contributes novel extensions of existing SGDLM methodology to address this. 

 The canonical setting is that the $q$ series are observed over time prior to an identified event-- or intervention-- time $T$. At this time, a subset of the series is regarded as potentially affected, with the rest assumed to be unaffected. The setting will often be observational though we follow prior literature in using \lq\lq experimental'' and \lq\lq control'' for the two sets of series.  Reorder the series so that  $\y_t'=(\y_{ct}',\y_{et}')$ with $q_c-$vector $\y_{ct}$ of controls and $q_e-$vector $\y_{et}$ of experimental series for
$q_c+q_e=q.$  The same terminology applies for $t<T$.  
The counterfactual forecasting focus is to predict  $(\y_{et}|\y_{ct})$ for $t>T$ and then assess how such predictions depart from the post$-T$ development of $\y_{ct}.$ over that period.    The literature is embracing the more global goal of using all control series together, rather than constructing inspired summaries of them as synthetic controls~\citep{TierneyEtAl2024,TierneyEtAl2024Supplement,KevinLiEtAlCausalMVTS2024}. 

A key challenge in the multivariate context is the flexibility of modeling dependencies among series; traditional models can be overly constraining, especially for larger series dimension $q$.  It is important to enable sensitive characterisation of pre-intervention dependencies between experimental and control series; counterfactual forecasting is wholly based on that.  We also emphasise routine sequential analysis in which the post-intervention inferences monitor potential causal changes, so multivariate time series models that enable sequential analysis are of interest. This is particularly important in terms of learning about cross-series relationships that may change in time.  Further, the ability to define a parallel post-intervention analysis that explicitly admits the potential for changes-- providing an alternative to the counterfactual hypothesis  in the context of evolving, sequential learning from post-intervention data, is a key theme.  This relates to a number of interests, including the question of whether some of the putative control series do in fact exhibit post-intervention changes potentially related to the intervention, i.e., so-called interference effects. The SGDLM framework provides opportunity to address these desiderata.

\subsection{Counterfactual Analysis in SGDLMs\label{sec:CFManalysis}} 

Given a pre-intervention SGDLM fitted over $t=1:T,$  counterfactual analysis begins under the assumption of no intervention effects. Use CFM to denote this  counterfactual model. Write $\y_{e_0t}$ for  the purely hypothetical counterfactual outcomes for $t\ge T,$ data that would have occurred under this assumption. Pre-intervention,  $\y_{ct}$ and $\y_{e_0t}\equiv\y_{et}$ are observed; post-intervention the CFM assumes only $\y_{ct}$ is observed and $\y_{e_0t}$ is \lq\lq missing data''.   Predictions of $\y_{e_0t}$  are made and successively revised over the post-intervention period to compare with the actual  outcomes; denote the latter by $\y_{e_1t}.$    

CFM analysis requires extension of the existing SGDLM methodology since the missingness of $\y_{e_0t}$ complicates post-intervention inference: 
the conjugate normal-gamma updates for the univariate models cannot be performed if experimental series are either parents or children of other series. 
This can be addressed by incorporating the missing data $\y_{e_0t}$, together with the parameters $\bTheta_t,\bLambda_t$, in a novel Monte Carlo filtering strategy that neatly exploits the model structure. 

The target filtered posterior for $\bTheta_t,\bLambda_t,\y_{e_0t}$ at time $t$ has the compositional form
$$p(\bTheta_t,\bLambda_t|\y_{ct},\y_{e_0t},\cD_{t-1}) 
p(\y_{e_0t}|\y_{ct},\cD_{t-1}).$$ The first term is the standard SGDLM posterior for states and volatilities  assuming $\y_{e_0t}$ known.  The second term, the marginal posterior for the missing data,  is not available analytically but can be accurately approximated as follows. 
Given the time $t-1$ prior Monte Carlo sample of states and volatilities, 
$\bTheta_t^r,\bLambda_t^r$,  $(r=\seq 1R),$  
the joint time $t-1$ p.d.f. of $\y_{ct},\y_{e_0t}$ has direct Monte Carlo approximation given by
 $$ p(\y_{ct},\y_{e_0t}|\cD_{t-1}) = R^{-1}\sum_{r=\seq 1R} 
   p(\y_{ct},\y_{e_0t}|\bTheta_t^r,\bLambda_t^r,\cD_{t-1}) $$
 where each term is a multivariate normal p.d.f. from the SGDLM. 
 The implied conditional for the missing data $\y_{e_0t}$ given the observed $\y_{ct}$ is then
\begin{equation}\label{eq:MCmixpy0givenyc} 
p(\y_{e_0t}|\y_{ct},\cD_{t-1}) = \\ \sum_{r=\seq 1R}  \pi_{rt}(\y_{ct}) \ 
         p(\y_{e_0t}|\y_{ct},\bTheta_t^r,\bLambda_t^r,\cD_{t-1})
\end{equation}
 where the component p.d.f.s are the implied conditional normals for $\y_{e_0t}$, and the mixture weights $\pi_{rt}(\y_{ct})  \propto p(\y_{ct}|\bTheta_t^r,\bLambda_t^r,\cD_{t-1})$ simply involve evaluations of the corresponding marginal normal p.d.f.s at $\y_{ct}.$ 
The strategy is then to generate a sample of $\y_{e_0t}$ from the above mixture of normals, and on each sampled value generate $\bTheta_t,\bLambda_t|\y_{ct},\y_{e_0t},\cD_{t-1}$ with the usual SGDLM importance sampling weights proportional to $|\I-\bGamma_t|_+.$   The resulting IS weights will reflect additional variation due to uncertainty about $\y_{e_0t}$; this is implicit as each parameter sample relies on its corresponding sampled value of the missing data. 
 
Computations involve multivariate normal density evaluations and multinomial sampling.  Technical details and   relevant theoretical developments are given in Appendix~\ref{app:smcdetails}.
The MC sample size $R$ can be increased with limited computational demands, and this may be important in improving the resulting IS performance.

\subsection{Outcome Adaptive Model}  An important methodological development in Bayesian causal analysis, introduced by~\cite{KevinLiEtAlCausalMVTS2024} using other time series models, is to also consider a separate {\em outcome adaptive model} (OAM). 
In the current setting, the OAM is the same SGDLM used for the CFM analysis,  and is run independently of, and in parallel to, the CFM.  Prior to 
intervention time $T$ the two are the same.  The difference is that the OAM
analysis updates post-intervention treating the $\y_{e_1t}$ as observed data along with the $\y_{ct}$; then at and only at time $T-1,$ 
the univariate DLMs for each of the {\em experimental series only} are adjusted to allow for potentially higher levels of change in states and volatilities.
 After updating at  time $T$, the model reverts to precisely as pre-intervention.  The mechanism is to simply reduce the state and volatility discount factors at $T-1$ then revert to their normal, higher values at time $T$.  This is a standard method of responding to the possibility of a causal impact on the $e_1$ series, allowing the OAM to adapt to, and infer,  resulting changes on the explicit assumption that \lq\lq something may have happened''~\citep[e.g.][chap.~11]{West1986,WestHarrison1997}.  
 
 Causal prediction based on comparing inferences from the OAM  with those from the CFM provides a formal statistical assessment of the effects.   The OAM provides a global alternative (\lq\lq maybe something happened'') to the CFM null hypothesis (\lq\lq no change due to intervention'').
 As a side-benefit, exploring post-intervention predictions for each of the designated control series can prove useful in assessing potential interference effects, i.e., to what extent one or more controls may show inheritance of any changes seen in the $e_1$ series.

\newpage
  
\section{GDP Example: Illustrating Graphical and Factor Structure\label{sec:application}}

\subsection{Setting and Background} 

The series are annual real per capita GDP (1960--2003) in \$US for a selection of $q=16$ OECD countries, taken from \cite{abadie2015comparative}.  The countries with standard $3-$character codes are
\nar{Germany  (DEU), 
Austria (AUT), 
Denmark  (DNK), 
Netherlands (NLD), 
Belgium (BEL), 
France  (FRA),      
Italy (ITA),    
Spain (ESP),
Portugal (PRT),  
Switzerland (CHE), 
Norway (NOR),   
United Kingdom (GBR),        
Japan (JPN),
United States of America (USA),  
Australia  (AUS) and 
New Zealand (NZD), 
}
\noindent Here DEU is West Germany prior to the 1990 reunification with East Germany. 
A key interest is assessing the impact of that reunification on GDP of Germany and other countries. 
\begin{figure}[b!]
\centering
\includegraphics[clip,width=0.7\columnwidth]{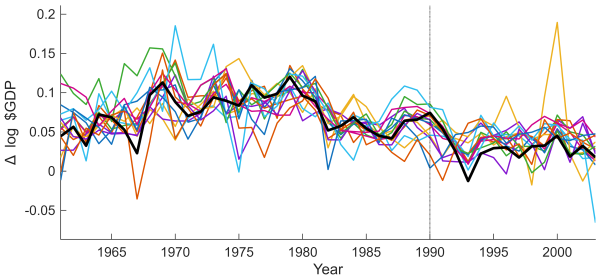}  
\caption{Changes in annual log GDP of $q=16$ countries. The thick line is DEU. 
  \label{fig:GDPlogchangeseries}}  
\end{figure} 
This data set has been used to exemplify other causal methodological developments~\citep[e.g.][]{klosner_comparative_2018,pang_bayesian_2022}. In contrast to these prior analyses, which analyze GDP directly, we model the statistically natural annual log returns $y_{jt} = $   $\log(\GDP_{jt}/\GDP_{j,t-1})$ for country $j$ in year $t$; see Figure~\ref{fig:GDPlogchangeseries}. Contextual predictions then simply 
transform to the GDP scale. 
Most importantly, we define a different set of control countries than other analyses. Recognizing the potential for effects of German reunification on economic growth in other countries-- especially those in Europe-- we designate only Australia and New Zealand as control series. The remaining 14 countries are designated as experimental series along with Germany. In addition to contextual relevance, this moves well beyond prior models in terms of larger numbers of experimental series.

\subsection{SGDLM Specification and Monte Carlo} 

The univariate DLM for each of the $j=1:16$ series has a $3-$vector $\x_{jt}$  with first element 1, followed by lag$-1$ outcomes of  Australia and New Zealand series.  This defines a time-varying, sparse vector autoregressive form for $\bmu_t$ to (partly) predict short-term changes based on the lagged control series, with local intercepts to capture otherwise unexplained trends.   This model form is overlaid with the following numerical specifications.

\smallskip\noindent{\bf Priors:}
For $j \in 1:q$, the specification at time $t = 1961$ has state means $\m_{j,1961}^* = (0.05,0,...,0)'$, variance matrices $\M_{j,1961}^* = \text{diag}(0.0025,0.1,...,0.1)$, and volatility parameters $n_{j,1961}^* = 4$ and $s_{j,1961}^* = 0.0004$. The prior mean of each intercept is 0.05, representing  $\sim{5\%}$ GDP growth, while prior means of all other coefficients are 0. The overall priors are relatively diffuse in terms of both prior variances and a low initial degrees of freedom. The data over the first several years will therefore lead to substantial adaptation in the filtering analysis. 

\smallskip\noindent{\bf Discount factors:} Two state discount factors are used, one for  $\bphi_{jt}$ and one for $\bgamma_{jt}$ in~\eqn{SGDLMj}, with 
 the standard block discount strategy for DLMs~\citep[][sect.~6.3]{WestHarrison1997}. These are $\delta = 0.95$ for both state discount factors as well as $\beta = 0.95$ for volatilities. These allow for around 5\% increased uncertainty in the evolution of state and volatility parameters each year. 
 
\smallskip\noindent{\bf OAM intervention discount factor:} 
The OAM analysis sets $\delta=0.50$ at (and only at)  $t=1990$ in order that the model be more adaptive to immediate post-intervention data; the OAM analysis reverts to using $\delta = 0.95$ again from $t=1991$ onwards.  
 
\smallskip\noindent{\bf Monte Carlo:}
Analysis is based on $R=10{,}000$ Monte Carlo samples for SGDLM importance sampling steps each time period. Across  analyses with multiple SGDLM specifications, monitored values of 
effective sample sizes exceed 85\% for all years, indicating high efficiency~\citep[c.f.][]{GruberWest2016,GruberWest2017}  over the years.  This is true both pre- and post-intervention at $t=1990$.  Note that the post-intervention CFM analysis explicitly addresses unobserved counterfactual outcomes on all of the 14 designated experimental series; the high importance sampling efficiency is nevertheless maintained.    Direct simulations of posteriors for the \lq\lq missing'' experimental outcomes in the CFM also use $R=10{,}000$ samples each year.  Analysis summaries  below are based on Monte Carlo samples of filtered posteriors over time for state vectors and volatilities, and corresponding predictive distributions for future outcomes including the counterfactuals of causal interest.  
 
\subsection{Parental and Graphical Structure: An Example\label{sec:nicegraphexample}} 
 
To exemplify graphical and factor structure, a chosen set of parental predictors $sp(i)$, $(i=\seq 1q),$ is shown in Figure~\ref{fig:SGDLMgraph}. For each country $i,$ node $i$ in the graph has incident arrows from only those countries in $sp(i).$ 
This is one of a set of parental set specifications used in the more formal analysis-- Section~\ref{sec:GDPexampleBMA} below-- that allows uncertainty over parental sets.  It is, of course, important to stress that this and all other directed graphs evaluated in no way represent causal relationships (or, in this case, economic \lq\lq theories''); they are purely based  on the statistical evaluation of model:data match. Marginal likelihood evaluations yield this example graph as one of the more highly supported, sparse structures underlying this data in the SGDLM framework.    

Figure~\ref{fig:sparseGamma1990OAM} displays part of the sparse structure in each $\bGamma_t$ (white indicates zeros;  series order is irrelevant to the analysis and chosen only 
for nice visuals in the factor model structure implied, below). 
With this parental graph, each $\bGamma_t$ has 7 zero columns corresponding to the 7 series that are not parental predictors of any other, and those columns are omitted in the figure, for clarity. 
 The numerical values of the non-zero $\gamma_{ijt}$ superimposed are on-line posterior means from the OAM filtering analysis at time $t=1990.$  For example, the regression prediction for DEU at that point  has a simultaneous parental contribution of 0.46 BEL$+$0.35 USA, a weighted sum of 1990 GDP from the two parental predictors indicated in the graph.   The sparsity pattern induced in each precision matrix $\bOmega_t$ is shown in a 0/1 image in Figure~\ref{fig:GDPgraphprecision} 

\begin{figure}[htbp!]
    \centering
     \includegraphics[clip,width=0.55\columnwidth]{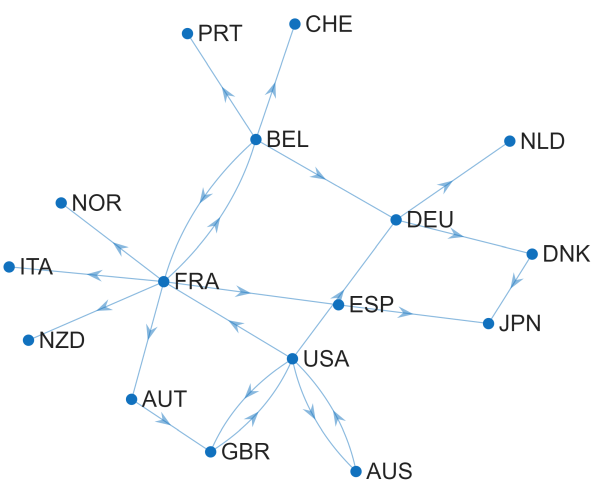}
    \caption{An example GDP simultaneous parental graph \label{fig:SGDLMgraph}}
\end{figure}

\begin{figure}[htbp!]
\centering
\includegraphics[clip,width=.45\columnwidth]{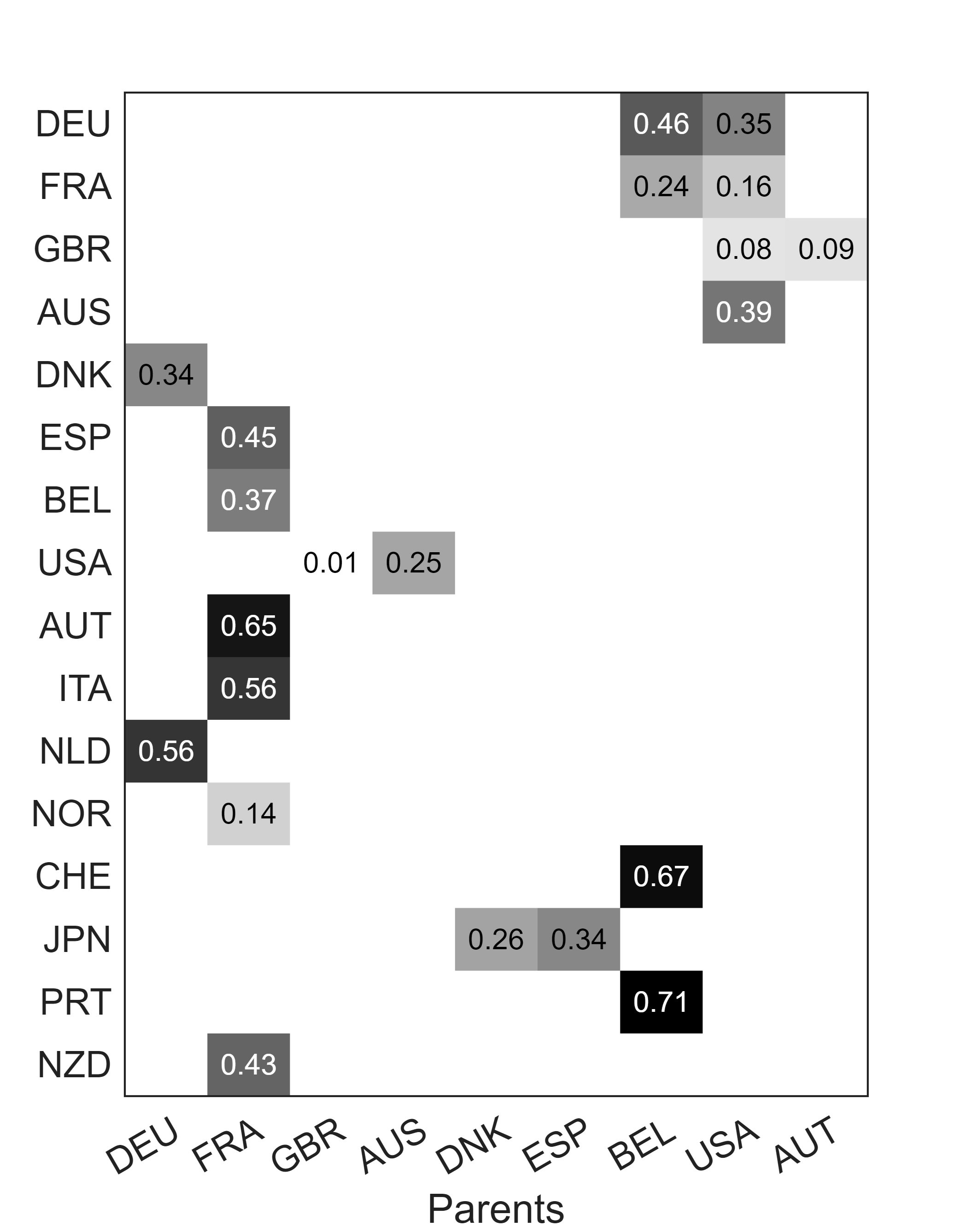} 
\caption{GDP example: Sparsity pattern in $\bGamma_t$  with OAM-based posterior means of non-zero values at $t=1990$. 
\label{fig:sparseGamma1990OAM}}
\end{figure}

\begin{figure}[htbp!]
\centering
\includegraphics[clip,width=.55\columnwidth]{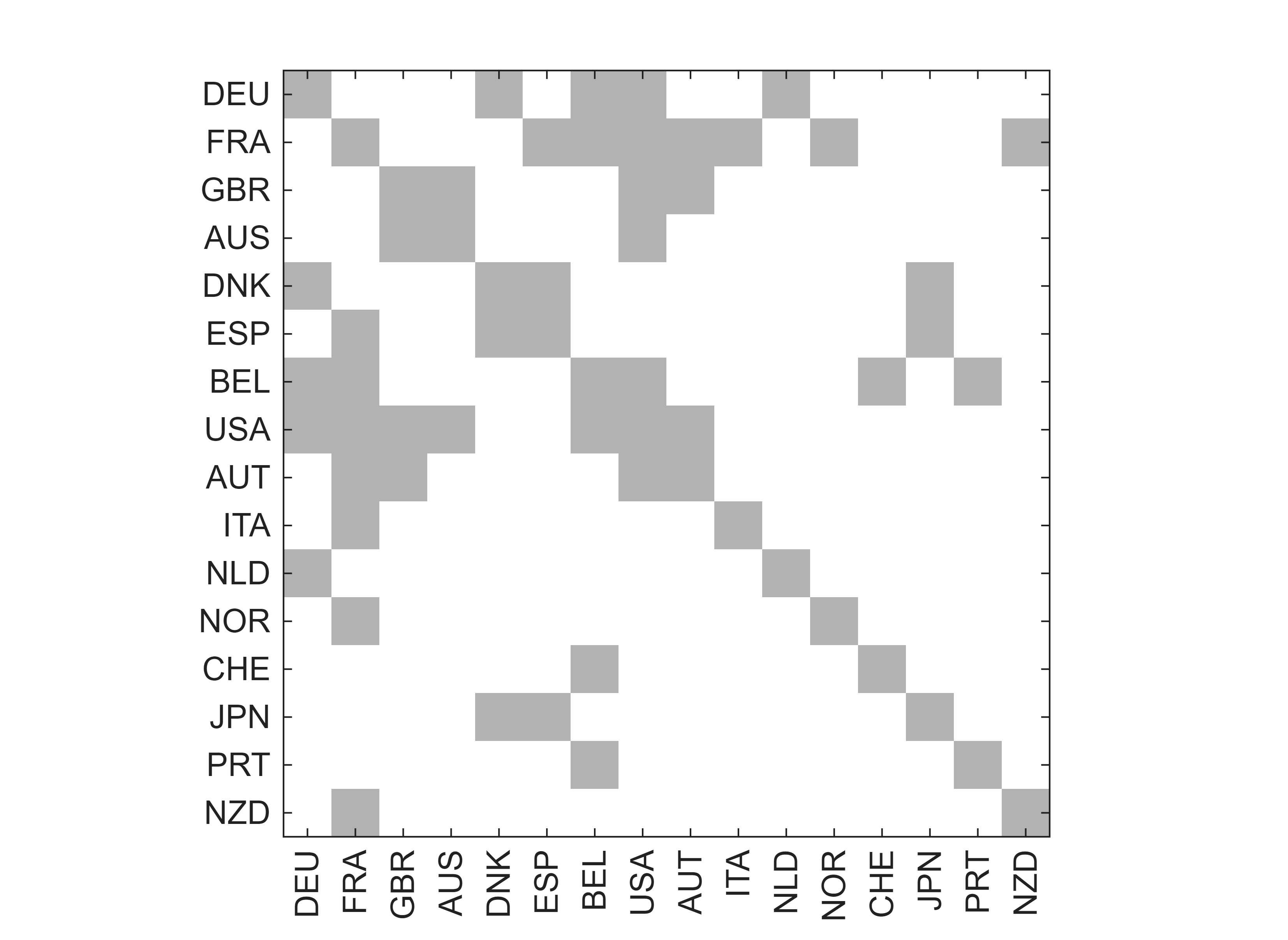}
\caption{Sparsity pattern of precision matrices $\bOmega_t$  implied by the GDP parental structure of Figure~\ref{fig:SGDLMgraph}.
\label{fig:GDPgraphprecision}}
\end{figure}

Of theoretical interest in the general graphical modeling area~\citep[e.g.][]{Jones2005a, Jones2005}, note that: (i) 
the directed simultaneous parental graph in Figure~\ref{fig:SGDLMgraph} has cycles; and (ii)  the implied conditional independence graph underlying  Figure~\ref{fig:GDPgraphprecision} is non-decomposable (it could be made decomposable by either the deletion of a single edge or the addition of two edges). 

\FloatBarrier

 \subsection{Example of Sparse Dynamic Factor Structure\label{sec:GDPsparsityandfactors}} 
 
Continuing with the example model of Section~\ref{sec:nicegraphexample},
 selected summaries of the implied dynamic factor structure of Section~\ref{sec:dynamicfactors} provide further insights. 
In each year $t$, inference on $\bGamma_t$ maps to that on factor loadings, singular values and 
scores $\fbA_t,\fD_t,\fbF_t$, respectively, and the implied factor process $\fbphi_t$.    This applies the SVD-based mapping in each of the Monte Carlo samples representing the current filtered posterior in each year; this of course includes samples of the missing $\y_{e_0t}$ in the post-intervention period. There are minor technical details associated with the inherent ambiguities in factor signs, ordering and matching between time points; see Appendix~\ref{app:factorID}.

\parablu{Graph-Factor Structure.} 
While the non-zero elements in $\bGamma_t, \fbA_t,\fbF_t$ are time-varying, the SGDLM graph defines fixed patterns of sparsity;  this aids in interpretation of the factors.  In this example, there are 5 common parental sets.
 Also, $\textrm{rank}(\bGamma)=7$ implying $p=7$ factors based on the set of  unique parental predictors; 
some of the 5 parental sets underlie more than one factor, discussed further below.

\begin{figure}[t!]
\centering
\includegraphics[clip,width=.45\columnwidth]{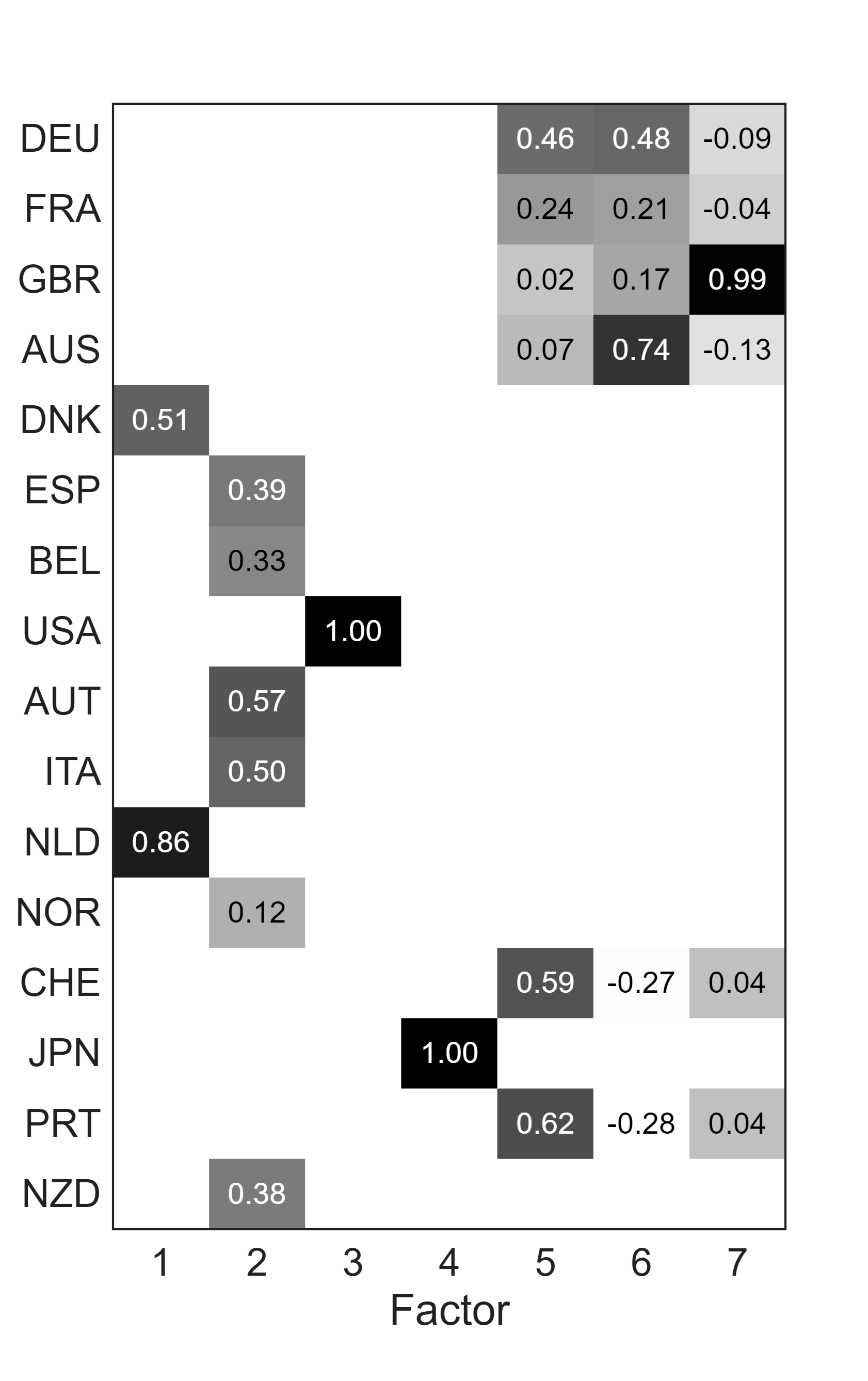}\\
\vskip-.15in\footnotesize Loadings matrix $\fbA_t$: $t=1990$\\ \vskip.075in
\includegraphics[clip,width=.45\columnwidth]{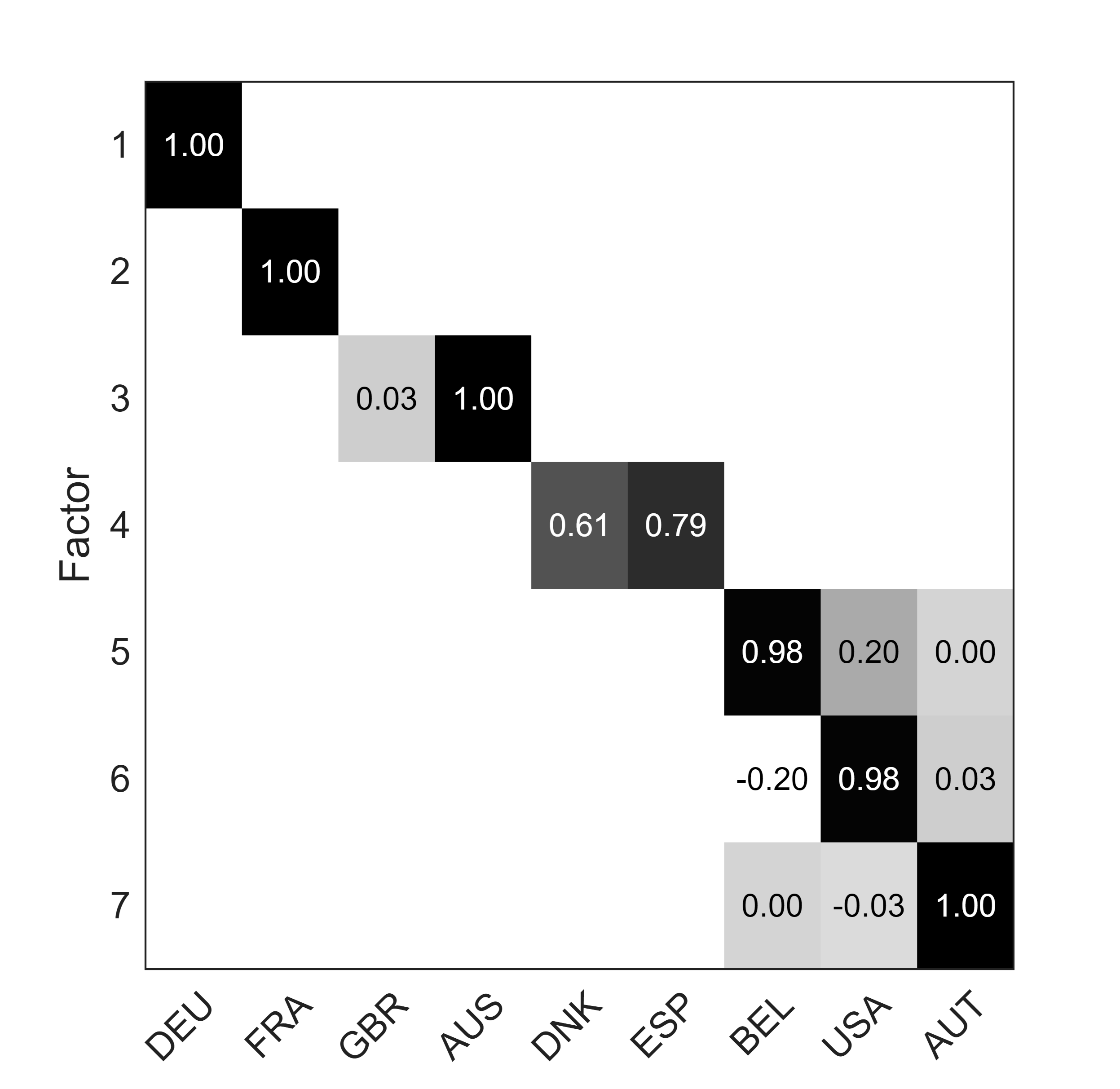} \\
\footnotesize Scores matrix $\fbF_t$: $t=1990$
\caption{GDP example: Sparsity patterns and factor structure in $\fbA_t$ and $\fbF_t$ with OAM-based posterior means of non-zero values at $t=1990$. 
\label{fig:sparseLS1990OAM}}
\end{figure}  

Figure~\ref{fig:sparseLS1990OAM} displays the sparsity of factor structure with numerical summaries providing a 
snapshots of OAM-based posteriors at $t=1990$.  In the displays of $\fbA_t$ and $\fbF_t$, white indicates zero while
shaded cells are labeled with on-line posterior means of the implied factor loadings and scores, respectively.  
The zero/non-zero factor pattern in $\fbF_t$ relates to the 5 common parental sets $\cP_h$, $(h=\seq 15)$.  
With factors in the (arbitrary) order shown:   
(1)  $\cP_1=\{\textrm{DEU}\}$ alone defines factor 1; 
(2)  $\cP_2=\{\textrm{FRA}\}$ alone defines factor 2; 
(3)  $\cP_3=\{\textrm{GBR, AUS}\}$ defines factor 3;
(d) $ \cP_4=\{\textrm{DNK, ESP}\}$ defines factor 4, and
(3)  $\cP_5=\{\textrm{BEL, USA, AUT}\}$ represents series inter-dependencies reflected in 
 factors 5,6 and 7. 
  
The factor structure-- in terms of the zeros/non-zeros-- is fixed over time, though inferences on the non-zero values evolve in the sequential analysis. Corresponding figures for other years show a relatively high level of stability of many non-zero values, though there are some marked differences. Some of the main apparent changes come over 
  1991-1993, immediately following German reunification, when the GDP dropped in DEU and most other countries.
These are related to the impacts of the reunification intervention but overlaid and confounded by the developing recession over those years across the Western economies, in particular, related to a range of additional, potentially causal contributors.

\parablu{GDP Dynamic Factors.}  
Inference on factor structure includes the on-line, filtered posterior Monte Carlo samples for the underlying latent factor process~$\fphi.$ Figure~\ref{fig:GDPfactors1457mcOAM} gives examples, showing trajectories of  4 selected factors.  

\begin{figure}[tp!]
\centering
\includegraphics[clip,width=0.7\columnwidth]{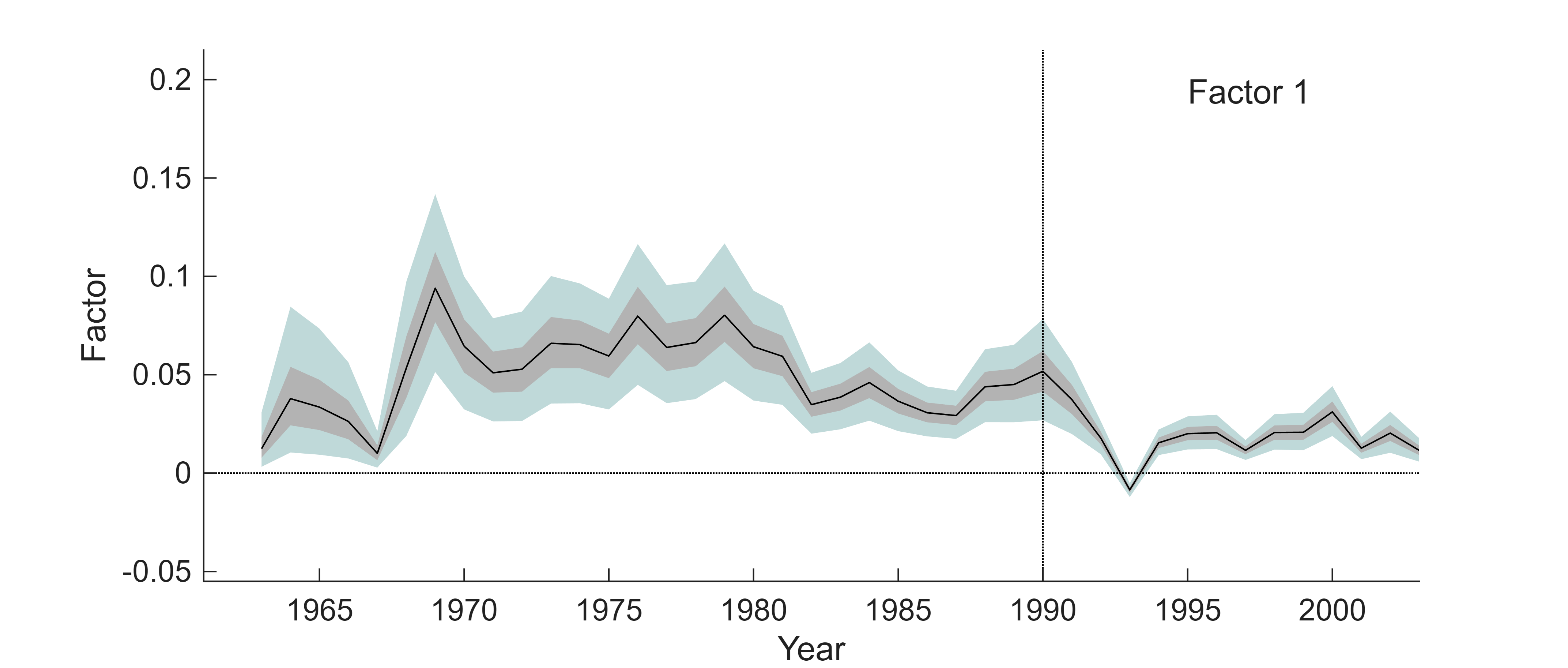} 
\includegraphics[clip,width=0.7\columnwidth]{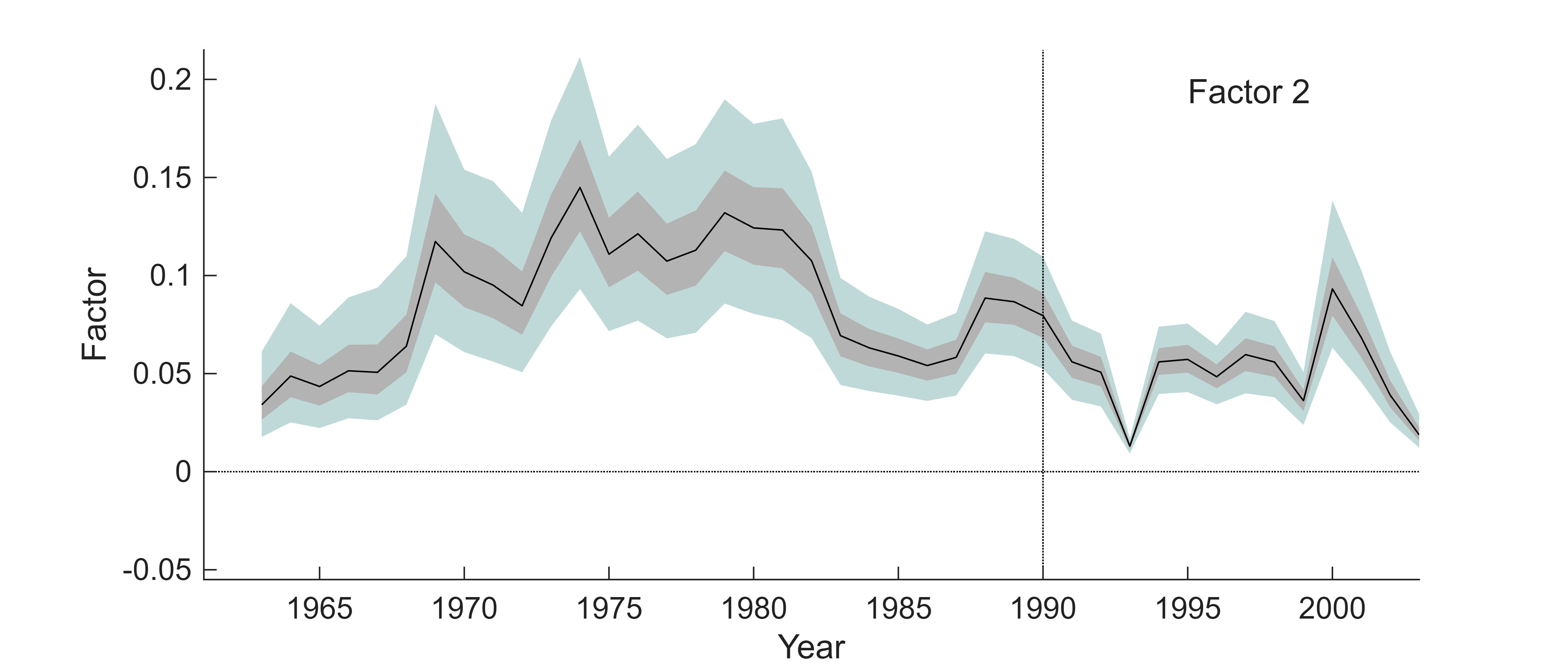}   
\includegraphics[clip,width=0.7\columnwidth]{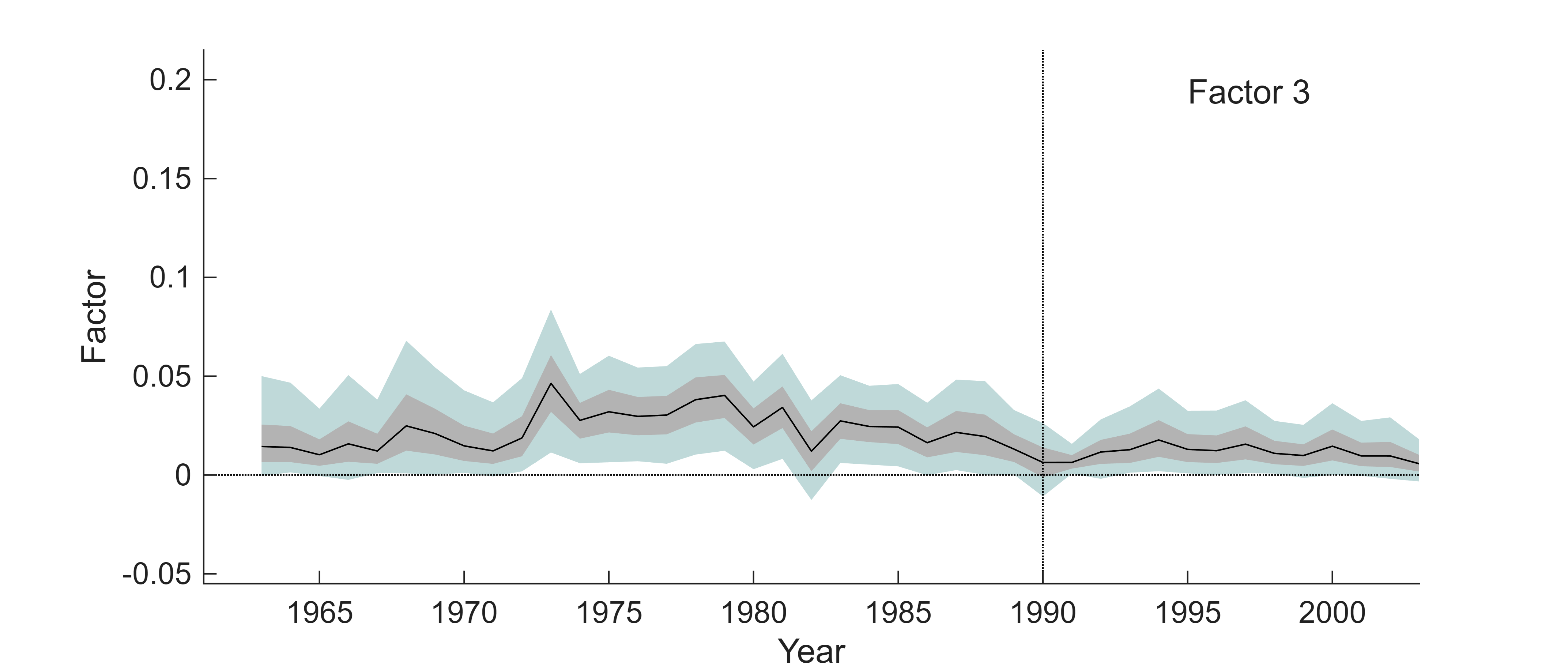} 
\includegraphics[clip,width=0.7\columnwidth]{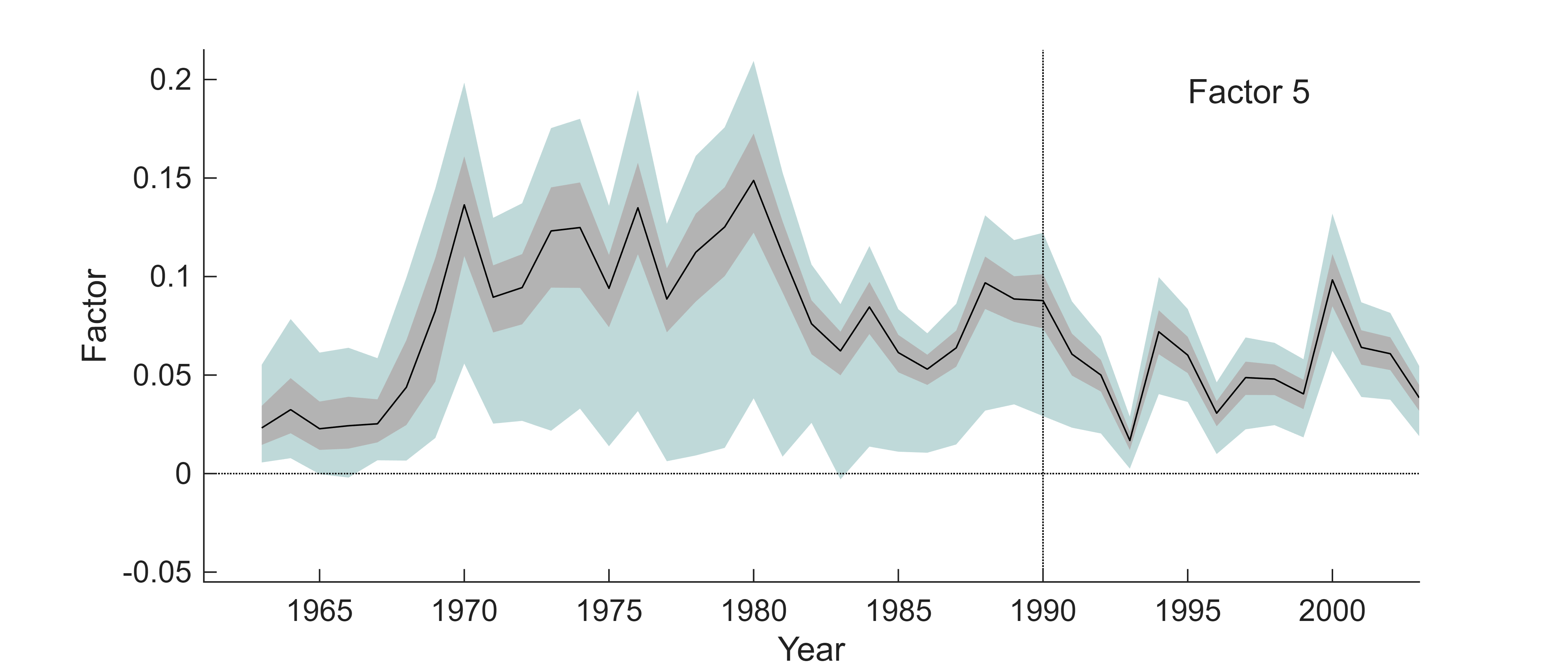} 
\caption{GDP example: OAM-based trajectories of filtered posteriors for factors, showing medians (full lines), and 
50/90\% (dark/light shading) equal-tails credible intervals. }
  \label{fig:GDPfactors1457mcOAM}
\end{figure}

\begin{figure}[pt!]
\centering
\includegraphics[clip,width=0.7\columnwidth]{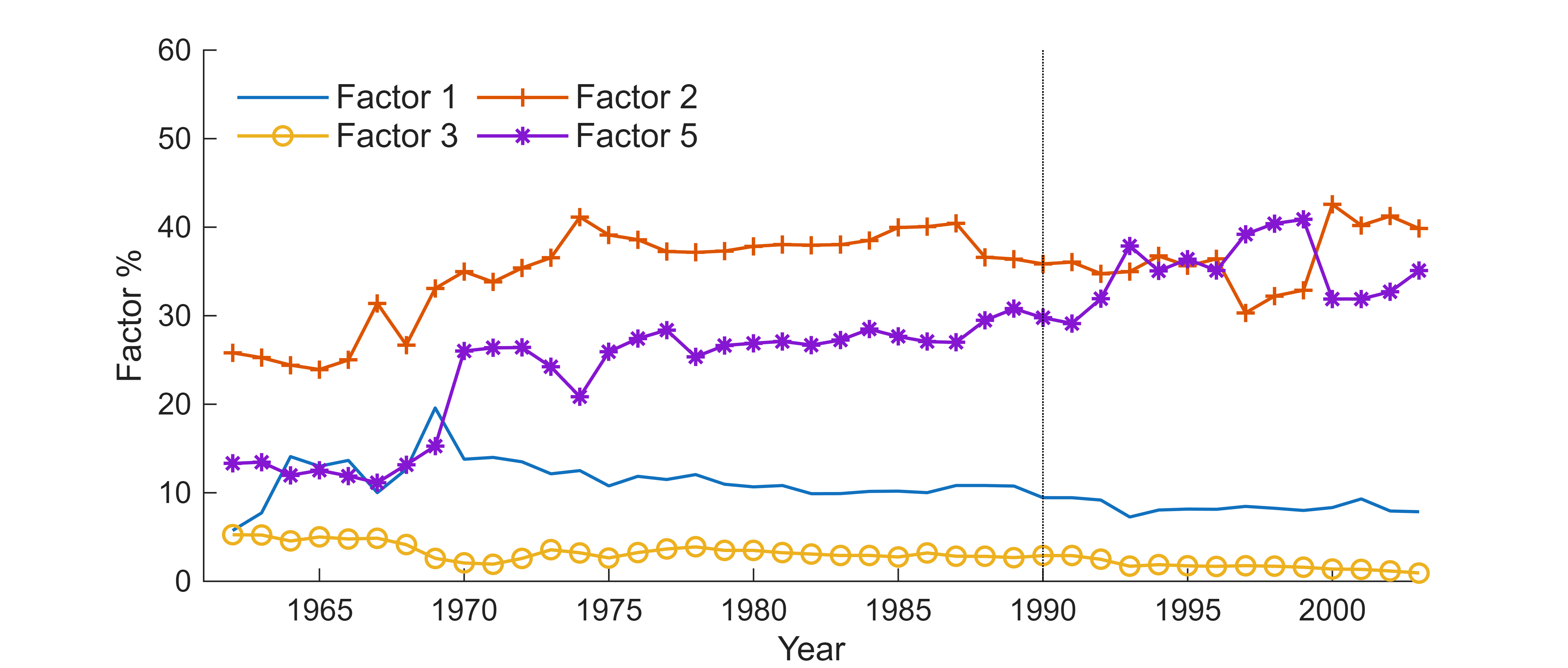}  \\
\caption{GDP example: OAM-based trajectories of   filtered posterior medians of \%-variation explained by 4 factors.  
  \label{fig:GDPfactorspercent}} 
\end{figure}
 
The initial several years represent a learning phase as the SGDLM analysis adapts to incoming data, but once into the early 1970s the nature of the factors becomes clear as accruing data informs on state vectors and volatilities.  
 The 4 factors chosen here highlight interpretation in the macroeconomic setting.  
 In addition to the behavior of DEU-founded factor 1 over 1991--1993,  some of the other factors show distinct drops over that period.  There is inherent contextual confounding of the impact of German reunification with other global economy events in the early 1990s that contributed to Western economy recessions, and the full set of factors reflect complex cross-series relationships arising as a result over those years.  
Factor~2 is the FRA-founded factor,  showing a post-intervention drop but then rebounding more than Factor~1. 
Factor~3 is the $\{\textrm{GBR, AUS}\}$-founded factor, showing marked stability over the full time period and no serious evidence of impact of the 1990 reunification intervention.   
Factor~5 is one of the three  $\{\textrm{BEL, USA, AUT}\}$-founded factors, shows substantial volatility over the years and a general trajectory similar to that of Factor~2. 
One modest feature is the  large uptick in Factor~2 in 2000; this is the only factor on which NOR has a non-zero loading, and 2000 was a year of major growth in NOR GDP driven mainly by the major increases in global oil prices over 1999-2000 
 to the benefit of GDP in what was then one of the premier producers and suppliers of crude oil.   A  similar but smaller up-tick is seen in Factor~5,  showing the cascade of impact of the oil market changes more globally. 
 These examples also highlight the fact that SGDLM-based factor models will often exhibit meaningful cross-factor dependencies.   

Figure~\ref{fig:GDPfactorspercent} provides information on the relevance of factors over time, showing OAM-based filtered posterior trajectories of percent variation explained by each of these factors 1,2,3 and 5. This just shows posterior medians in each year.   Factors 2 and 5 are major contributors to the levels and patterns of simultaneous variation over time. with increased levels over the full time period.  Post-1990, Factor~5 plays more of a role than earlier, while the 2000 period noted above sees it drop by to about the 1990 level.  Factors 1 and 3 discussed contribute to the full simultaneous description of the 16 series at much lower, though generally stable levels up to 1990, then showing some modest decreases post-1990. 
The comparable CFM-based analysis  (graph not shown here) 
is, of course, precisely the same prior to 1990, following which the trajectories remain essentially constant at the 1990 levels.

\section{GDP: Counterfactual Forecasting and Model Averaging Analysis\label{sec:GDPexampleBMA}}

\subsection{Perspective and Model Set Selection}
Analysis admitting some level of graphical structure and hyper-parameter uncertainty involves running multiple chosen SGDLMs in parallel, and then updating the model-specific marginal likelihood each year following details in Section~\ref{sec:marglikcalc}. A baseline analysis with initially uniform model probabilities is typically of interest; then the normalized marginal likelihoods at each $t$ define current posterior probabilities over models, and inferences can be averaged in the traditional Bayesian model averaging (BMA) mode~\cite[e.g.][sect.~12.2]{WestHarrison1997}. 
This is exemplified here with a focus on uncertainty about the parental set specification. This can of course be applied to include model hyperparameters, such as discount factors, in routine application. 

The perspective is that the graphical structure defined by parental sets is of interest in capturing cross-series dependencies, but not otherwise inherently interesting. As in other applications such as portfolio decisions~\citep[e.g.][]{GruberWest2016,GruberWest2017}, large sets of different parental structures can yield very similar inferences and forecasts in the counterfactual setting. Exploring a small number of different structures that seem to yield different inferences, and that might be regarded a representative of different, empirically supported but otherwise similar models, is then the practical recommendation. 
 
The example here typifies the approach: (a)  visit and evaluate many possible parental set specifications based on training, pre-intervention data; (b) select a very small number of the most highly scoring models based on that analysis, aiming to include models that are structurally different-- in this context, differing in levels of sparsity.  Use this small set of selected models to run the SGDLM analysis pre- and post-intervention.  In the GDP example, screening of models in part (a)   visited all  $2^{15}$ univariate models of each $y_{jt}$ using all possible subsets of the remaining $\y_{-j,t}$, evaluating the approximate 
marginal likelihoods and using a binomial prior for the number of per-series parental predictors. The latter uses 
independent parental inclusion probability $k'/p,$ encoding an expected $k'$ parents.  For each $k'=\seq 14$, two graphs are delivered by choosing the highest- and second-highest–scoring parental sets per series out of all subsets. From among these, step (b) 
takes 8 candidate models that have the same priors, discount factors, and exogenous predictors, but different parental sets.  This yields graphs with different levels of sparsity, ranging from an average of 1.3 to 2.6 parents per series

\subsection{SGDLM Analysis}

Exploiting the marginal likelihood evaluations of Section~\ref{sec:marglikcalc}, SGDLM analysis on each models yields cumulative Bayesian model probabilities (under an initially uniform prior) whose trajectories are 
shown in Figure~\ref{fig:BMAweights}. During the first decade of the pre-intervention period, models with fewer parents perform relatively well, especially the more sparse $k'=1$ models,  The models have very similar point forecast accuracy during this time, but less sparse models have higher forecast uncertainty stemming from the larger number of parental coefficients.  As the years progress, however, information about these coefficients accumulates, and the denser models are able to better characterize multivariate volatility in the series. The $k'=4$ model does increasingly well throughout the 1970s and 1980s, in large part due to its larger parental sets on just one or two series.

\begin{figure}[b!]
\centering
\includegraphics[clip,width=0.7\columnwidth]{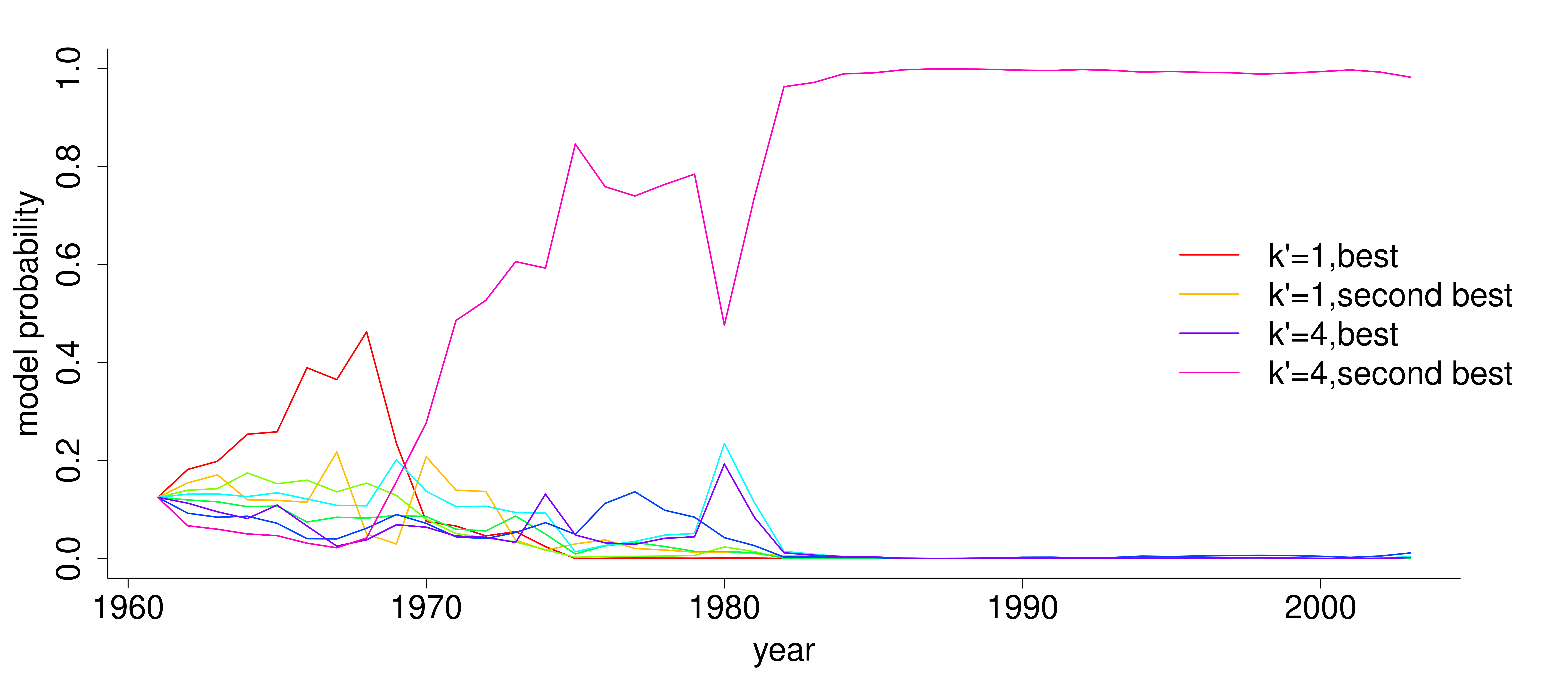}
\caption{BMA probability trajectories on the set of 8 models.
\label{fig:BMAweights}}
\end{figure}

The example of Section~\ref{sec:application} is based on the best scoring $k'=1$ model, i.e., that
 under the prior encouraging more sparsity in parental connections. That graphical structure serves well in illustrating the graph theory and factor model implications. The more complex model with $k'=4$ has richer structure that allows for more adequate reflection of increasingly minor aspects of cross-series structure and their changes over time. As noted above, the contributions from different series show that this is very heavily influenced by one or two series that are not significantly involved in the counterfactual analyses.  This is shown in reanalysis using only the single, sparse model with $k'=1$.  The results (not shown) define counterfactual forecasts and inferences that are practically indistinguishable and essentially similar to those of the full BMA analysis reported in the following sections. Other aspects of cross-series structure and dynamics will, of course, differ. This highlights a general point in model comparison and combination in multivariate settings. BMA (and other) analyses on the full set of series can favour models that-- while being statistically \lq\lq best" on the full data set-- can be of lesser interest in terms of specific inferential questions.

\subsection{Counterfactual Forecasting} 
Inferences are now based on averaging over models with respect to evaluated model probabilities representing uncertainties over parental set specifications. 
Figure~\ref{fig:counter_drawsGAB} shows filtered inferences on $\y_{e_0t}$ over time in the full SGDLM (pre-1990) and the CFM (post-1990), with outcomes superimposed, for Germany, Austria and UK.   Figure~\ref{fig:counter_drawsGAB_originalscale} transforms these to the original GDP scale.  Post-1990, all three countries have GDP growth below what would be expected under the counterfactual, but the departure from this trend is most striking in Germany, especially in 1992-1994. Based on the observed values of the controls, we would expect little change in German GDP growth in the first five years after reunification; instead, we observe a sharp drop-off. Uncertainty in the counterfactual forecasts naturally increases over time post-1990. 

\begin{figure}[t!]
\centering
\includegraphics[clip,width=0.7\columnwidth]{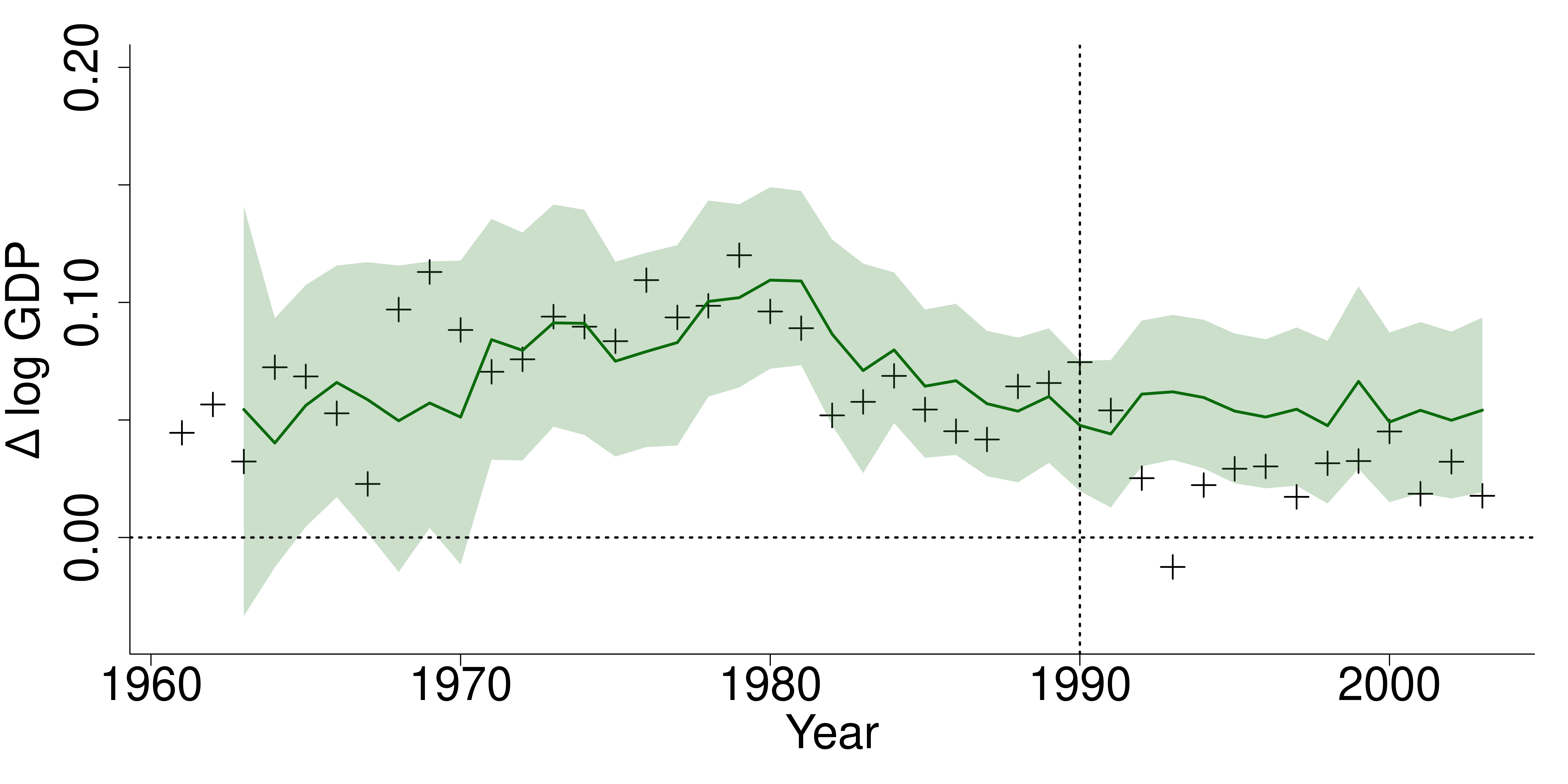} \\
\footnotesize (a) Germany \\
\includegraphics[clip,width=0.7\columnwidth]{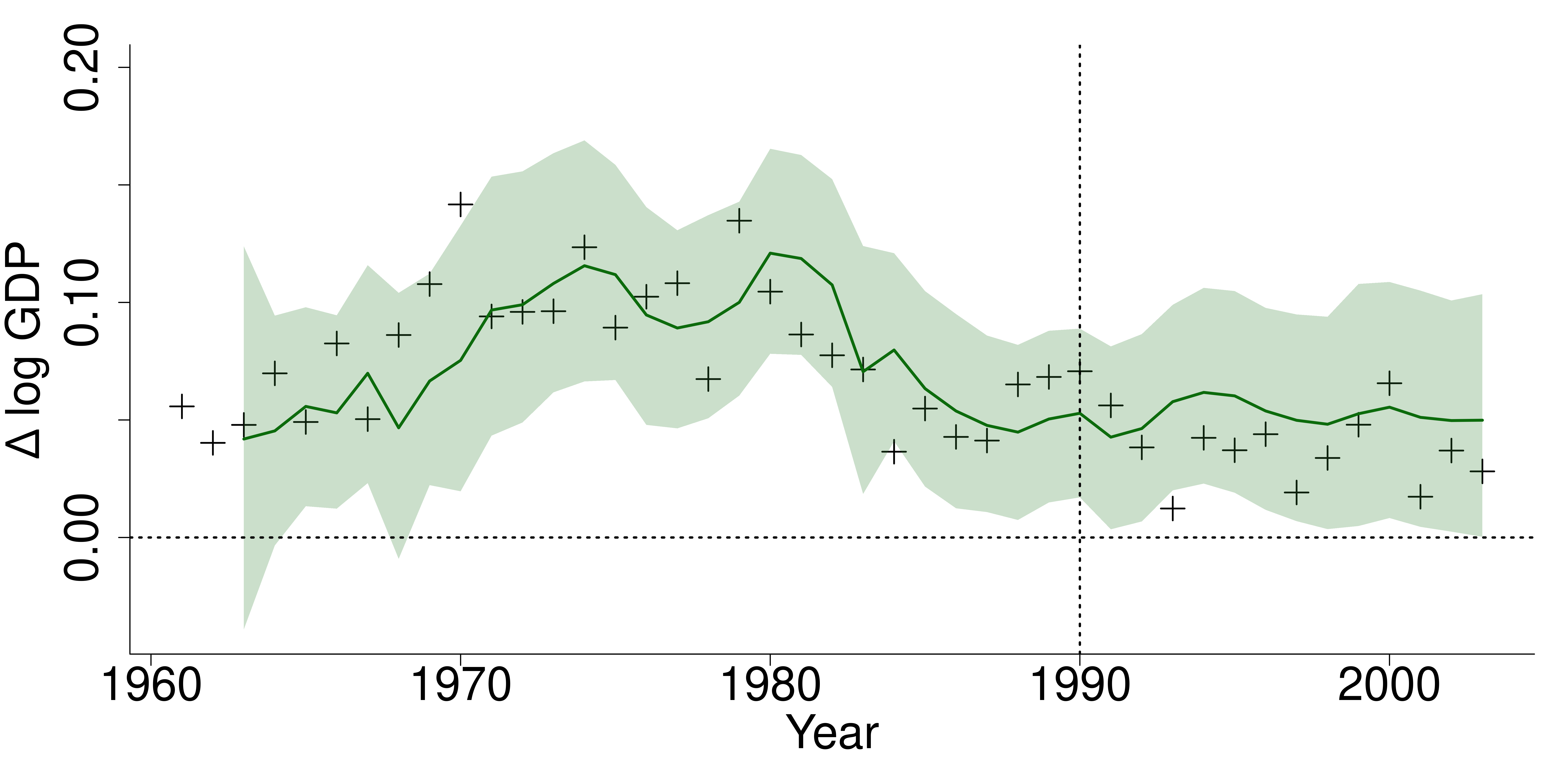} \\
\footnotesize (b) Austria \\
\includegraphics[clip,width=0.7\columnwidth]{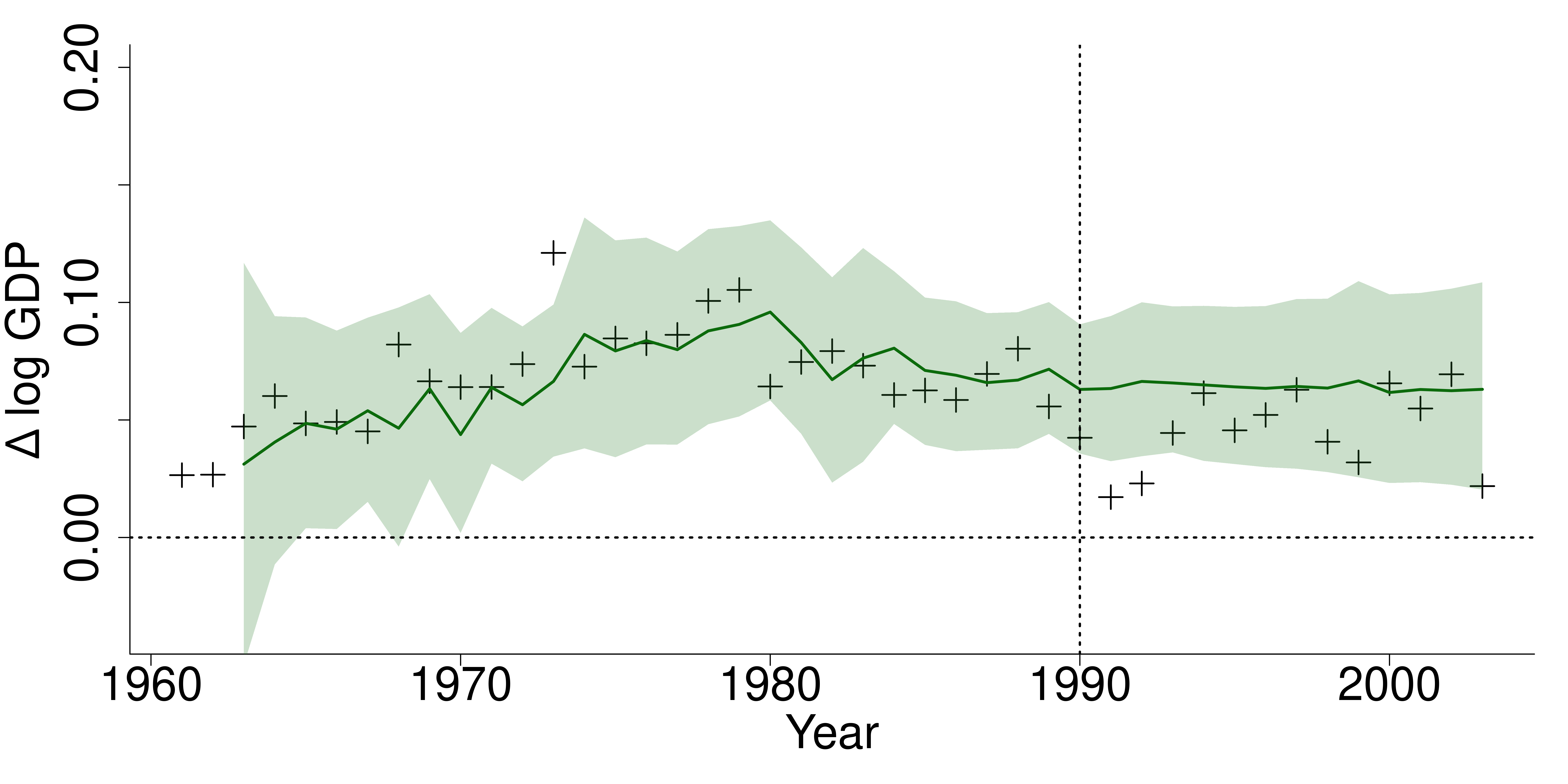} \\
\footnotesize (c) United Kingdom
\caption{Trajectories of nowcasts of elements of $\y_{e_0 t}$ for 3 countries, in terms of time $t$ filtered posterior medians (full line) and equal-tails 90\% intervals (shading),  with outcomes  (crosses) superimposed. At each $t$, nowcasts are from the current posterior for $\y_{e_0 t}$ conditional on $\y_{ct}$ and all past data. Pre-1990 the analysis updates fully to also condition on the observed  $\y_{et}$ before evolving to the next year. 
}
\label{fig:counter_drawsGAB}
\end{figure}

Figure~\ref{fig:filtered_meansGAB} shows inferred trajectories of the difference in model means $\F_{jt}' \btheta_{jt}$ between the CFM and OAM.
This highlights departure of beliefs about the expected level of GDP growth accounting for the inevitable year-to-year stochastic variation. The trajectories for all three countries favor
 negative values, most substantially for Germany immediately post-intervention.  For UK, the initially negative effect diminishes later in the post-intervention period. Along with other European countries not pictured, the effect for Austria favors negative values for most of the post-intervention period, although smaller in magnitude than that of Germany and with credible intervals generally including 0. These findings indicate a possible spillover effect of German reunification on series other countries.

\subsection{Formal CFM:OAM Comparison} 

Traditional methods of sequential Bayesian model monitoring~\citep[][Chapter 11]{WestHarrison1997} defines one summary numerical evaluation of 
the departure of the observed data from expectations under the CFM relative to the data-respecting OAM. This simply involves sequential updates of the relative model probabilities based on the ratio of time $t$  marginal likelihoods of Section~\ref{sec:marglikcalc}.  As the models differ only in the lower discount factor at $t=1990$, this provides an additional, quantitative assessment of the impact of the intervention.

Figure~\ref{fig:model_comparison} displays the resulting trajectory of the probability on the OAM relative to the CFM based on equal probabilities at 1989 (through which time the models are, of course,  the same). The increased uncertainty under the OAM just after the intervention leads to lower probabilities in 1990, but the evidence in favor of the OAM accumulates throughout the next several years.  The evidence becomes strong by 1993, when several European countries experienced recessionary effects with lower GDP and in which Germany had its only year of negative GDP growth from 1961 to 2003. These departures from past trends are not anticipated by the CFM. As a result, the probability on the OAM reaches 0.99 in 1994 and never falls below 0.99 thereafter. This large impact is heavily influenced by series-specific contributions from Germany, Spain and Portugal, in particular. Throughout the rest of the post-intervention period, only 1998 contributes positively in favor of the CFM.

\begin{figure}[t!]
\centering
\includegraphics[clip,width=0.7\columnwidth]{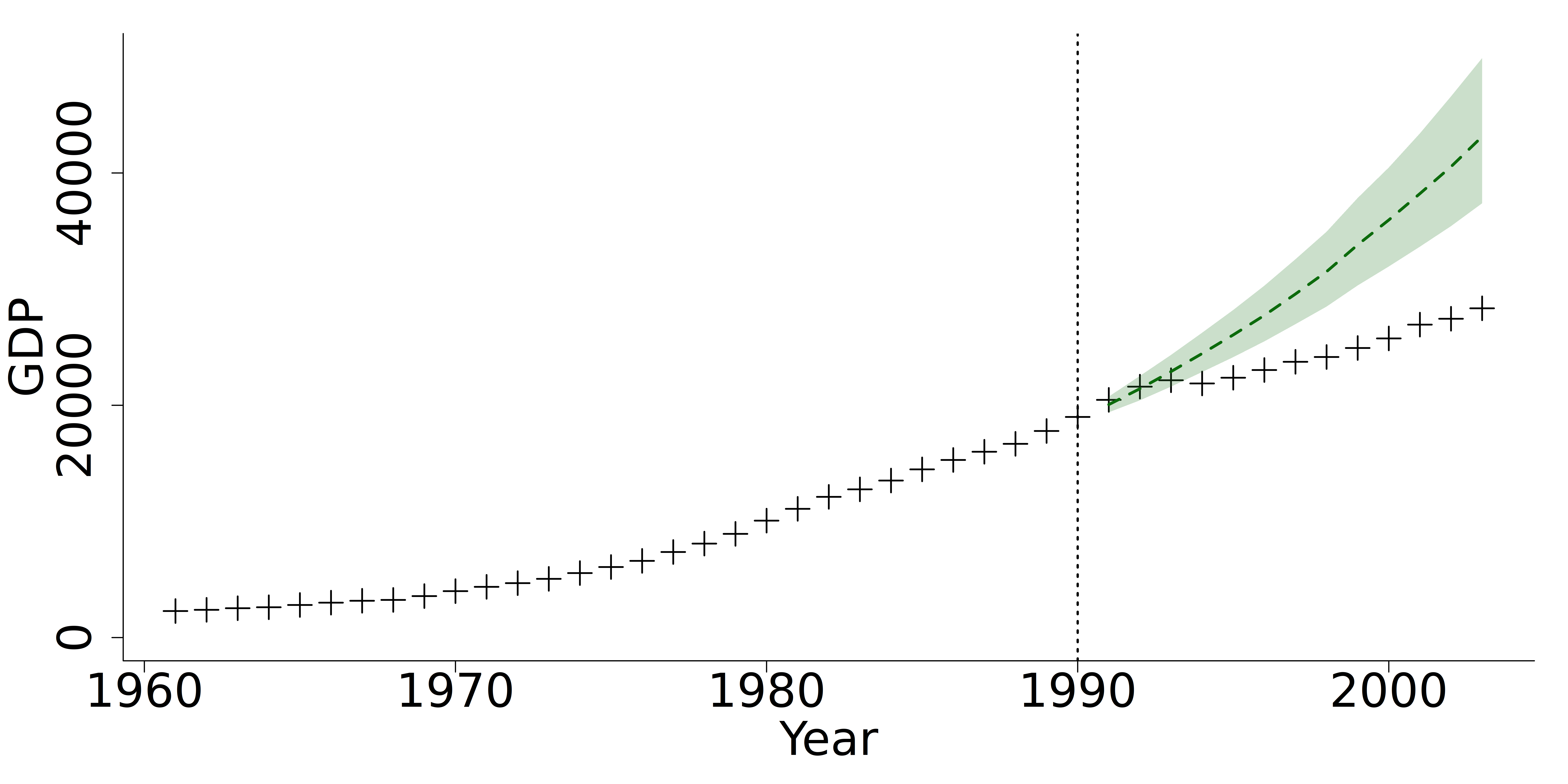} \\
\footnotesize (a) Germany \\
\includegraphics[clip,width=0.7\columnwidth]{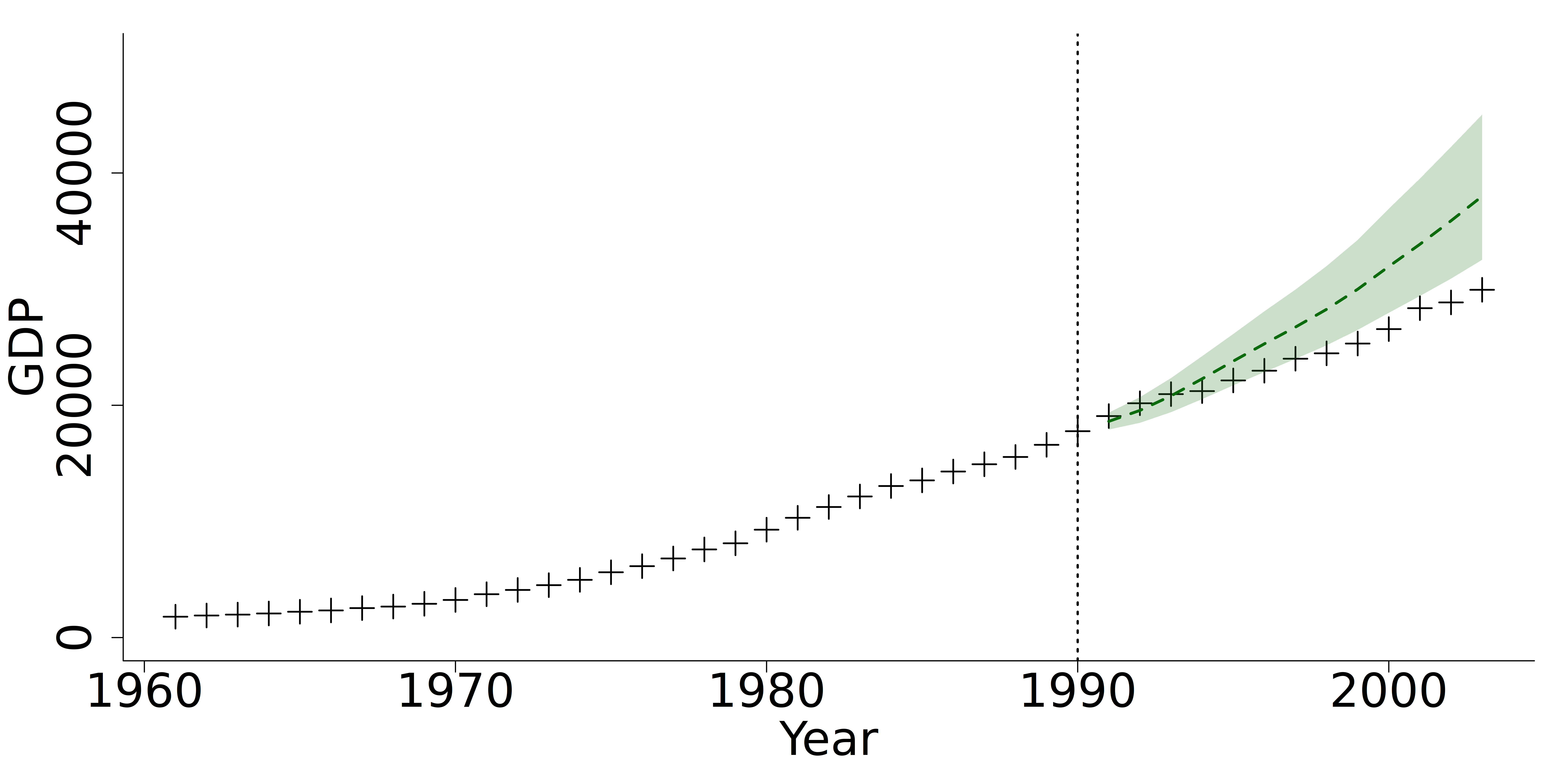} \\
\footnotesize (b) Austria \\
\includegraphics[clip,width=0.7\columnwidth]{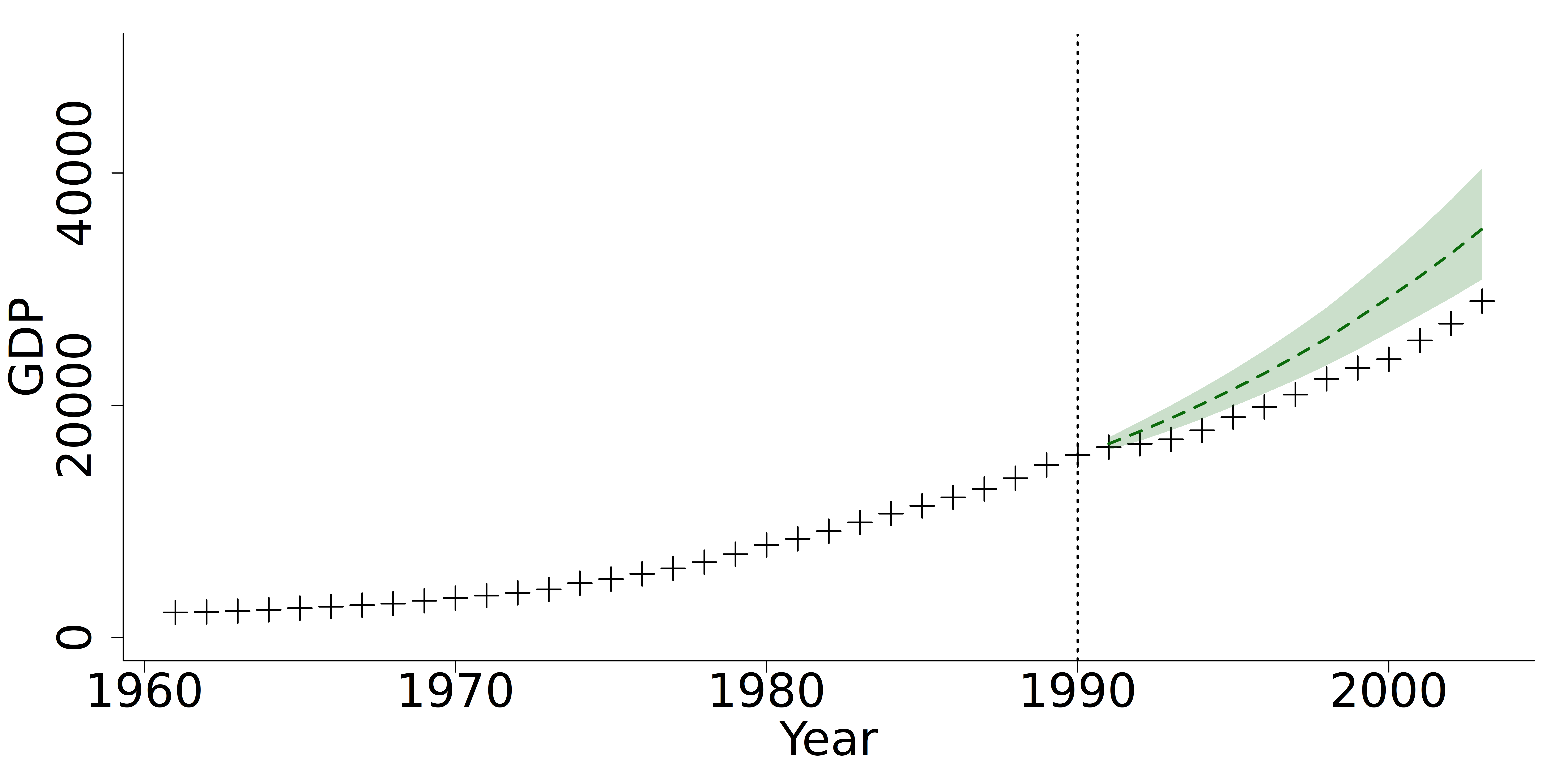}\\
\footnotesize (c) United Kingdom
\caption{Trajectories as in Figure~\ref{fig:counter_drawsGAB} transformed to the GDP scale.}
\label{fig:counter_drawsGAB_originalscale}
\end{figure}

\begin{figure}[t!]
\centering
\includegraphics[clip,width=0.7\columnwidth]{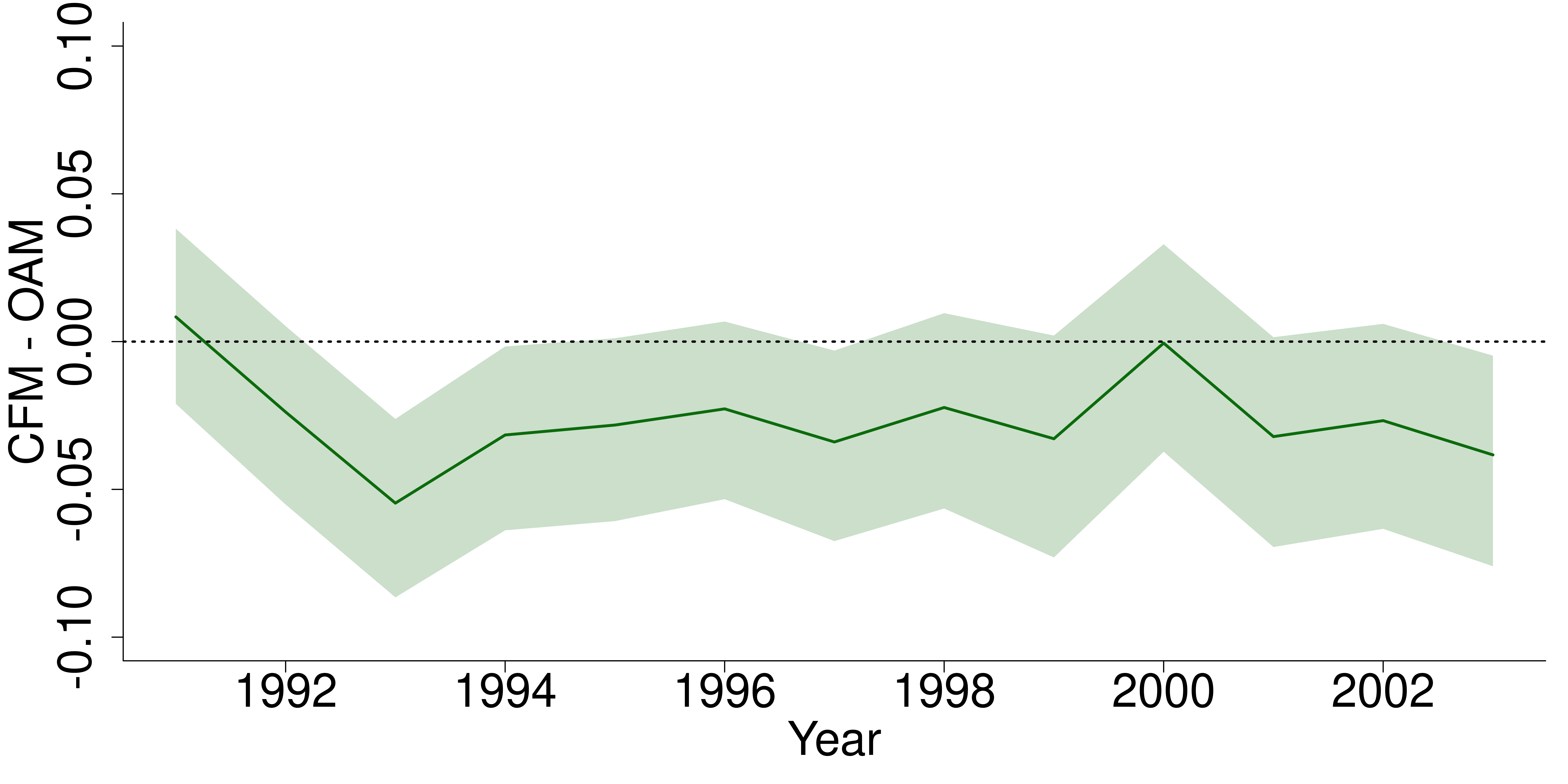} \\
\footnotesize (a) Germany \\
\includegraphics[clip,width=0.7\columnwidth]{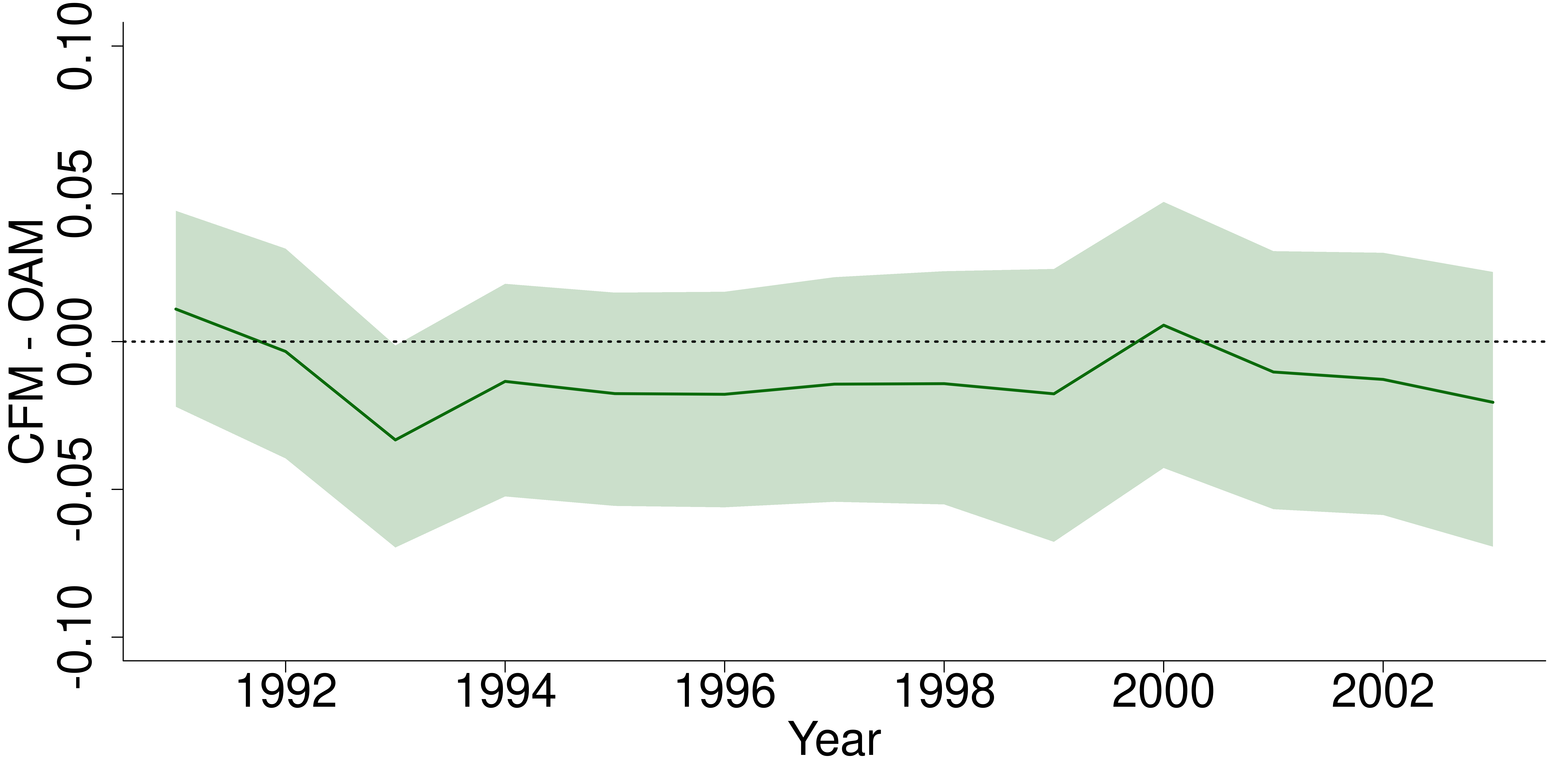} \\
\footnotesize (b) Austria \\
\includegraphics[clip,width=0.7\columnwidth]{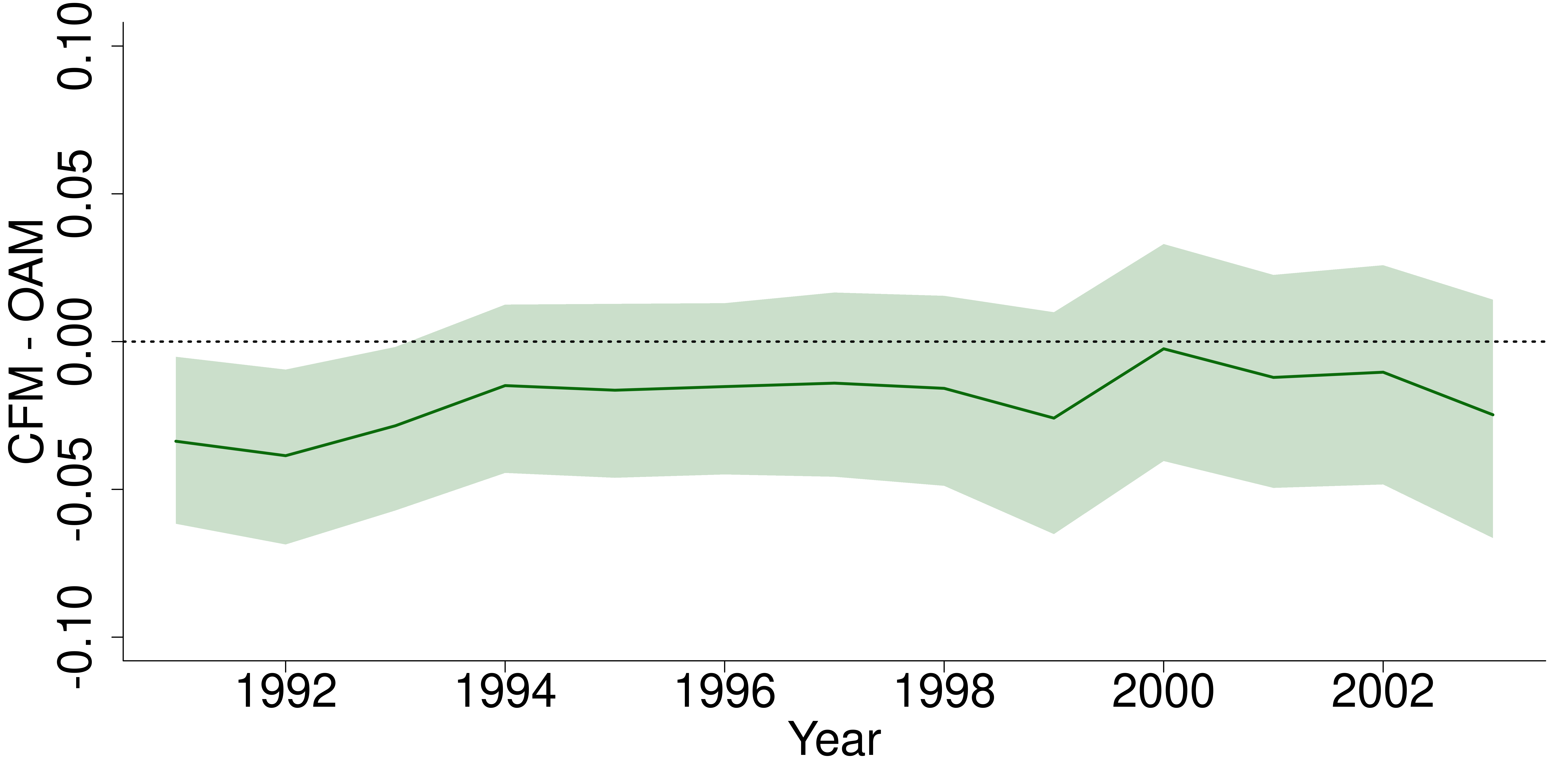} \\
\footnotesize (c) United Kingdom
\caption{Differences in filtered means: CFM minus OAM.}
\label{fig:filtered_meansGAB}
\end{figure}
 
\FloatBarrier
 
\begin{figure}[t!]
\centering
 \includegraphics[clip,width=0.7\columnwidth]{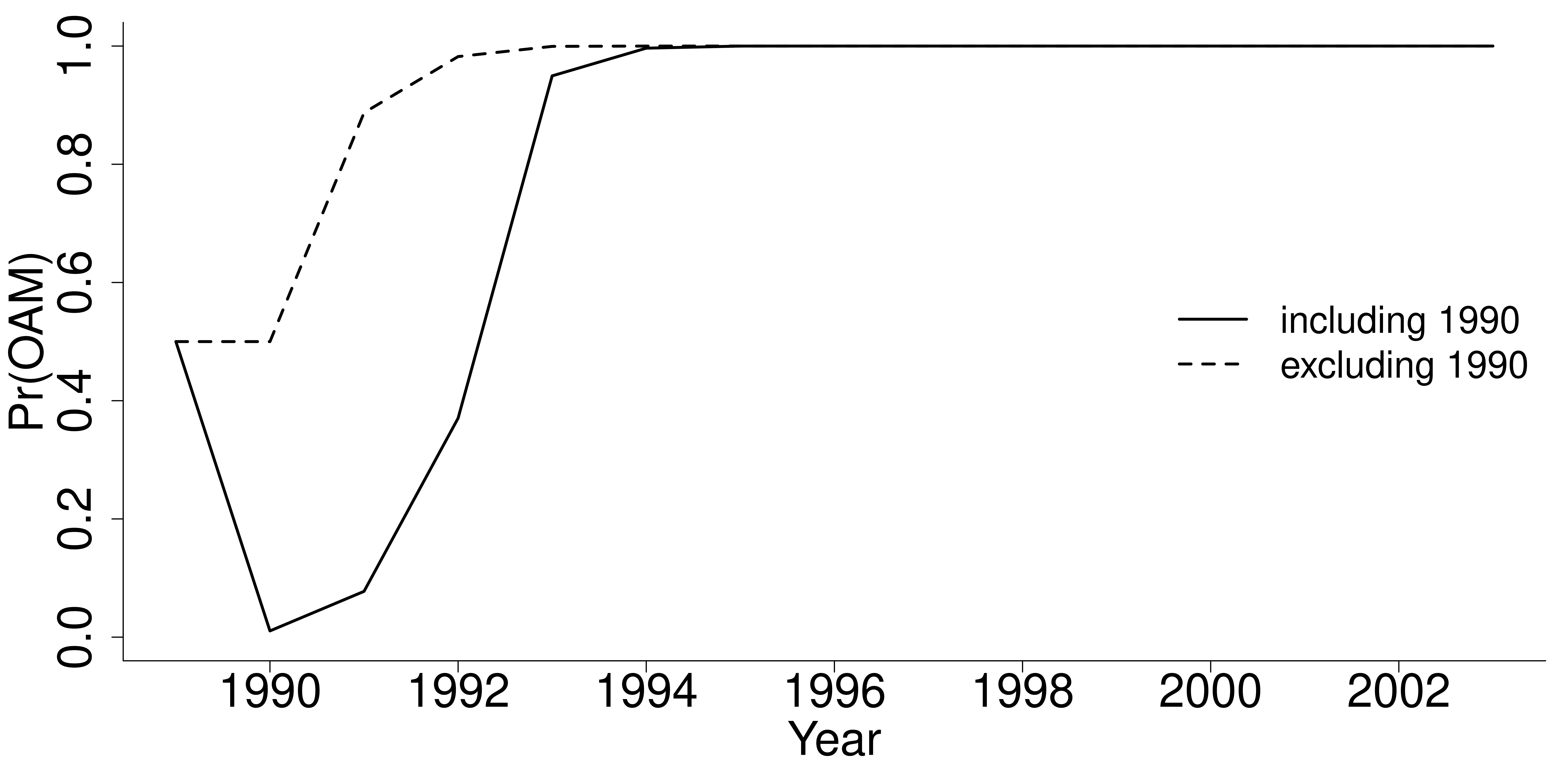} 
\caption{Post-1990 accumulating evidence in favor of the OAM over the CFM. 
\label{fig:model_comparison}}
\end{figure}


\subsection{Parental Regression Coefficients}

Time trajectories of posteriors for the $\gamma_{ijt}$ inform on changing cross-series relationships; this is of general interest while being more specifically relevant post-intervention. Figure~\ref{fig:gamma_germany_belgium_usa} displays this for coefficients of Germany on its two parental predictors Belgium and USA under the OAM analysis.  After a learning period the trajectories stabilize at positive levels up to 1990.  They then show immediate post-intervention changes with the simultaneous predictive role of Belgium increasing and that of the USA decreasing to negligible levels. 
The corresponding trajectories under the CFM (not displayed here) are the same as under the OAM pre-1990; after 1990 the median stays at approximately the 1990 level while the uncertainty slowly increases over the years.  
The OAM highlights and naturally adapts to this post-reunification shift, whereas the CFM naturally simply 
extrapolates the earlier, pre-intervention relationships with increasing uncertainty.

\begin{figure}[ht!]
\centering
\includegraphics[clip,width=0.7\columnwidth]{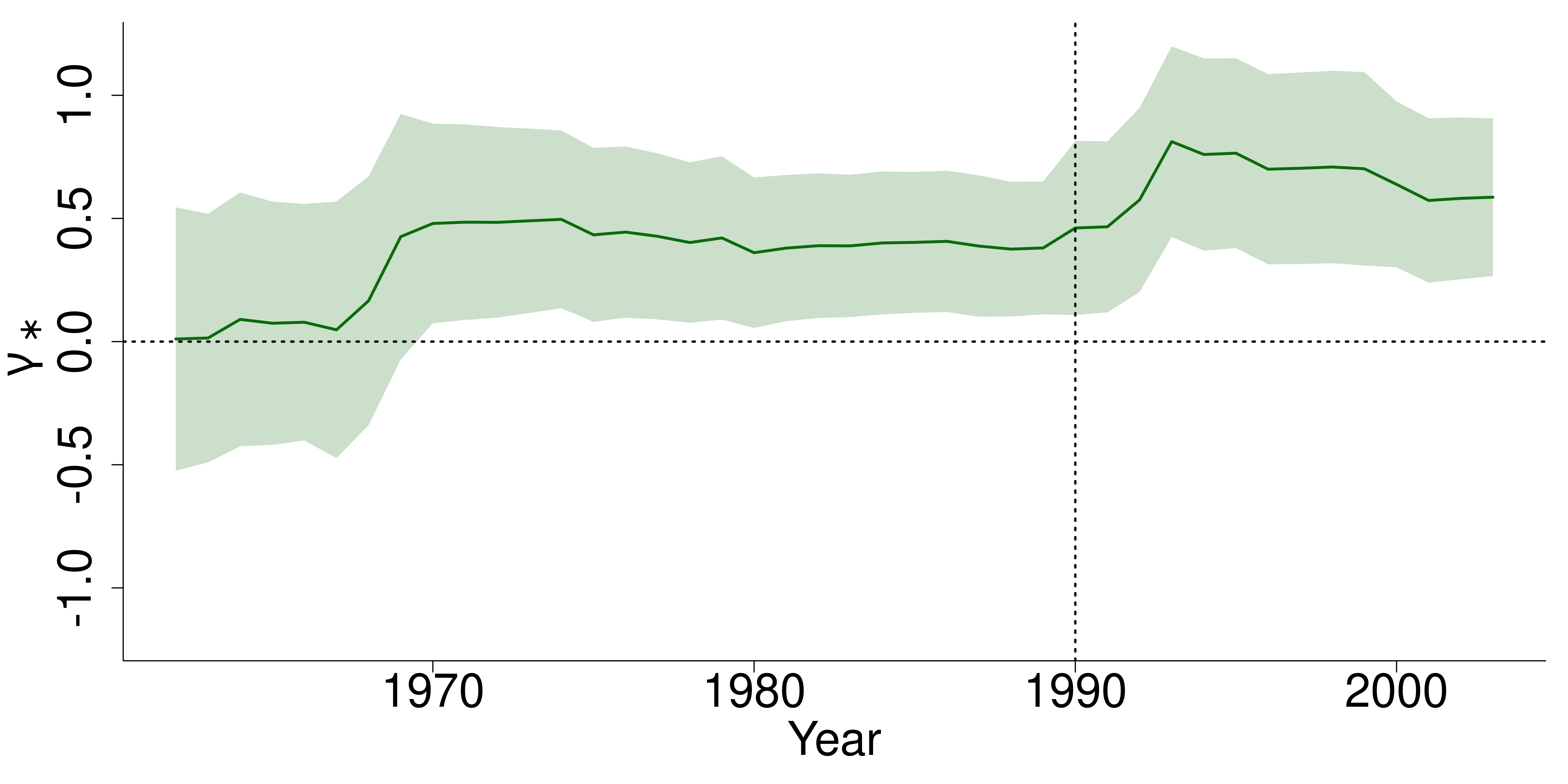}\\
\footnotesize (a) Germany on Belgium \\
\includegraphics[clip,width=0.7\columnwidth]{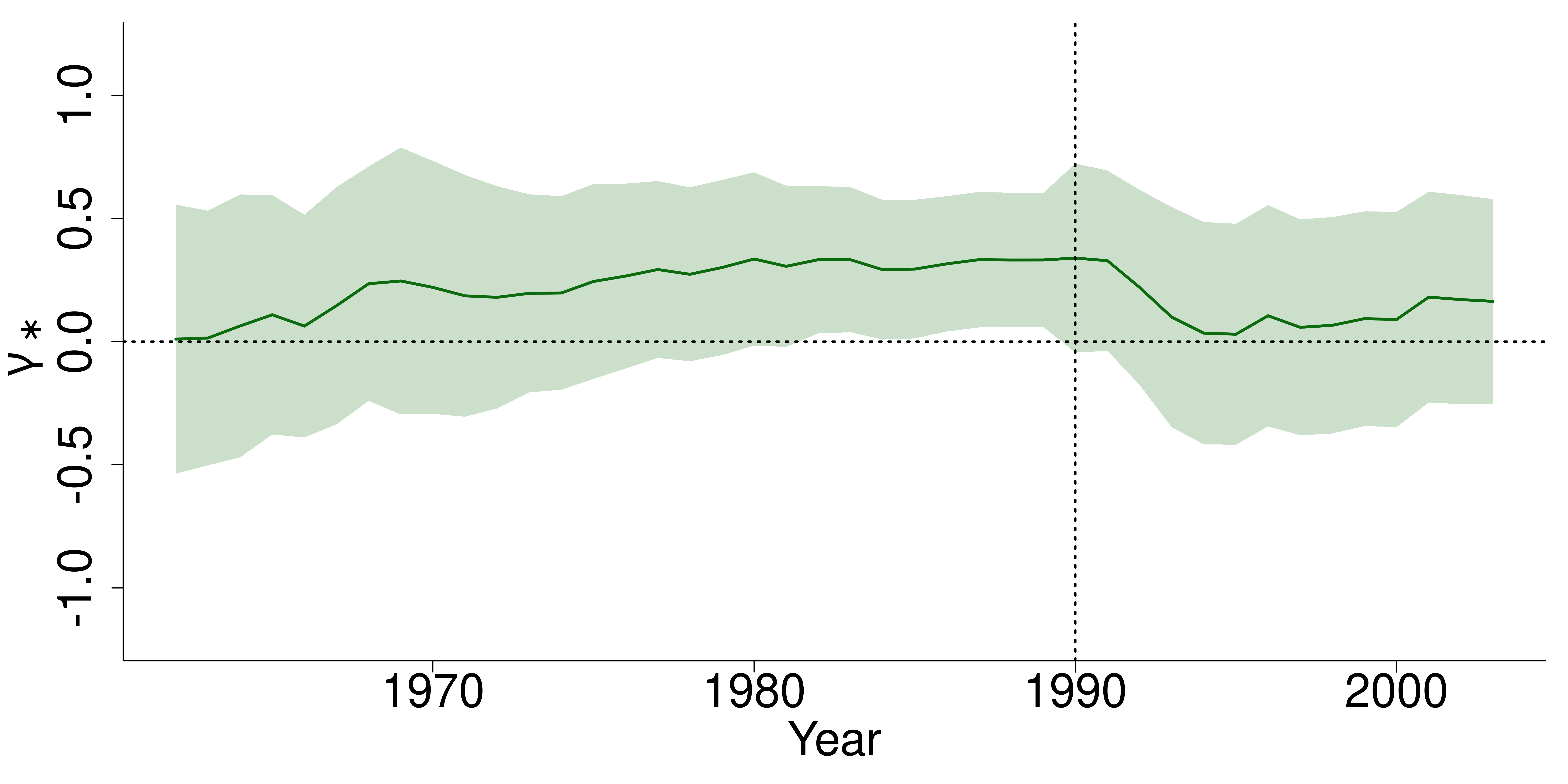} \\
\footnotesize (b) Germany on USA \\
\caption{OAM-based filtered trajectories for coefficients in $\bGamma_t$ of Germany on Belgium and USA.}
\label{fig:gamma_germany_belgium_usa}
\end{figure}

\section{Summary Comments\label{sec:summary}} 

 SGDLMs are increasingly used in applications such as forecasting for decision analysis in financial time series. Their broader utility is based on the inherent flexibility, parsimony and scalability with series dimension. Modelling flexibility is due to: (i) sensitive modeling of each of the decoupled univariate series separately,  enabling use of relevant exogenous and simultaneous predictors customised to each series; and (ii)  sensitive recoupling across the set of series through the simultaneous parental structure. This flexibility is enhanced by the inherent sequential model specification and Bayesian analysis that admits, evaluates and exploits time-variation in model parameters. Parsimony is enhanced by the decouple/recouple perspective and the use of underlying sparse graphical structure for cross-series relationships as part of that. Scalability is enabled-- computational loads scale linearly in the number of univariate time series-- as a result of the decouple/recouple modeling strategy. 

The graphical and factor model connections introduced here-- linking intimately to the graph theory underlying SGDLMs and related linear algebra-- define nice connections between what have been regarded as incompatible approaches to modeling multivariate dependencies. 
Sparse factor models have zeros in covariance matrices while sparse graphical models have zeros in precision matrices; these are generally incompatible~(though see~\citealp{Yoshida2010} for one prior existence proof of compatibility).  The general graphical structure of SGDLMs induces implicit factor structure, so providing an opening for new developments in factor modeling generally as well as specific to time series. 

Extension of SGDLM methodology to include the ability to evaluate marginal likelihoods across models addresses central practical questions of model uncertainty, and neatly exploits aspects of the mathematical structure of the existing SGDLM analysis. More major extensions to the 
setting of counterfactual forecasting is driven by putatively causal interests. This addresses questions of missing data on some of the univariate time series after a defined intervention time point, beyond which the designated experimental series are predicted based on the rest within the continuing evolution of the SGDLM. Bayesian computations-- relying only on simple, direct simulation-- remain sequential and efficient, exploit theory underlying the model structure, and neatly extend existing methods. 

\newpage

This paper has reviewed the core structure and main theoretical features of SGDLMs, and then reviewed and summarized the Bayesian methodology for sequential learning and forecasting in these models in the context of a given  parental structure. Exploring analyses against several graphs, chosen on empirical or contextual bases, is routine. Considering the graph as an underlying uncertain parameter defines research questions of dynamic variable selection and model uncertainty, and one for which the routine evaluation of model marginal likelihoods as they evolve over time is a central feature. In many applications,  identifying predictively useful graphical structures modulo specific goals is primary, and the extensions of SGDLM analysis here aid in that direction.  One aspect not entertained in the current paper is that of time-variation in the sparsity structure of models~\citep[e.g.][]{Jones2005a,Hans2007a,NakajimaWest2013JBES,LavineLindonWest2019avs}
 that are certainly of interest for potential future developments in counterfactual settings time series and allied areas are certainly of interest.  Some applications may involve  specific decision goals-- including potentially some applications in causal analysis where the inferences feed into actionable decisions.  In such settings, the parental/graphical structure selection question might best be embedded in model weightings relative to historical and near-term expected decision outcomes using ideas from Bayesian predictive decision synthesis~\citep{TallmanWest2023,TallmanWest2024}.

\begin{appendices}


\section*{Appendix} 

\section{Eigenstructure\label{app:eigenGamma}}
 
In Section~\ref{sec:ImpactofParents}, the matrix  $\I-\bGamma$ is invertible unless elements of $\bGamma$ are subject to pathological deterministic constraints. In the statistical setting with continuous priors and posteriors over the non-zero elements of $\bGamma$, invertibility holds with probability one since the eigenvalues of $\I-\bGamma$ will be non-zero with probability one. 

Write $\g=\{g_1,\ldots,g_q\}$ for the (real or complex) eigenvalues of $\bGamma.$  In applications, posteriors over $\bGamma$ will often favor regions where $|g_i|<1$ for all $i=1:q.$  The insights into spillover of Section~\ref{sec:ImpactofParents} are then implied. If, for example, the $y_i$ are all on the same scale, it is natural to expect that $\sum_{j\in sp(i)}|\gamma_{ij}|<1$  (i.e., expecting \lq\lq shrinkage to the mean'' in the univariate conditional models) which ensures all $|g_i|<1$ by the Gershgorin disc theorem~\citep[][Chapter 6]{HornJohnson2012}.  Empirical studies typically give estimated or simulated parameters with $\max(|g_i|)<1.$  Theoretically, however, some of the eigenvalues may lie outside the unit circle, but nothing changes in terms of the implied joint model. 
If desired, 
 Bayesian analyses of SGDLMs can be simply modified to reject cases in which one or more eigenvalues exceeds 1 in absolute value. This can be formally incorporated as a constraint in priors, and imposed in  weightings or accept/reject steps in the Monte Carlo (importance and/or MCMC) posterior analyses.  For example, 
 the reported GDP analyses used  $10{,}000$ Monte Carlo samples of $\bGamma_t$ over all 43 years $t$; the constraint  $\max(|g_i|)<1$ was satisfied by 99.98\% of the samples in both CFM and OAM analyses.

Some general points about the eigenstructure of $\bGamma$ follow from the corresponding directed graph. A \textit{cycle} is a path in the directed graph corresponding to $\bGamma$ that begins and ends at the same node, following the direction of the arrows, with an \textit{order} given by the number of nodes in the cycle. Let $r$ be the greatest number of nodes included in any collection of disjoint cycles in the directed graph.  The following results hold. 
\begin{itemize}
\item $\bGamma$ has at least $q-r$ zero eigenvalues, as $(-\delta)^{q-r}$ is a factor of the characteristic polynomial.

\item  For acyclic graphs, $r=0$ and $g_i=0$ for all $i = 1:q$~\citep[][Chapter 3]{cvetkovic_spectra_1980}. Then the order of the series can be permuted so that the resulting $\bGamma$ is strictly lower triangular with zero diagonal elements, and the simultaneous system reduces to a compositional factorization of $p(\y)$. 

\item If $\bGamma$ has at least one cycle and all cycles are of even order, then all non-zero eigenvalues of $\bGamma$ occur in pairs 
$\pm \lambda$~\citep[][Theorem 5]{marimont_system_1969}. When this condition holds for odd $q$, at least one of the $g_i$ is 0. 
\end{itemize}
    

\section{Marginal Likelihoods\label{app:marglikdetails}}

Evaluation of $1-$step ahead forecast density functions for marginal likelihoods are detailed here. For clarity here, the notation initially drops the $t$ index and conditioning  information set $\cD_{t-1}.$  Thus, in general form the p.d.f. of interest is
\begin{equation*} 
p(\y) =  
 \int |\I-\bGamma|_+ \prod_{j=1}^q p_j(y_j|\y_{sp(j)},\btheta_j,\lambda_j)p_j(\btheta_j,\lambda_j) d \bTheta d \bLambda
\end{equation*}
where the terms in the product are from normal linear regressions with conjugate NG priors. 
Marginalizing out the 
$\bphi_j,\lambda_j$ gives
\begin{equation} \label{eq:app:pygamma} 
p(\y) =  \int |\I-\bGamma|_+ \ \prod_{j=1}^q p_j(y_j|\y_{sp(j)},\bgamma_j)p_j(\bgamma_j) d \bGamma
\end{equation}
where $p_j(\bgamma_j)$ is the p.d.f. of the marginal multivariate T prior for the simultaneous parental subvector in series $j$.

\parablu{Naïve sampling from the prior $p(\bGamma)$}. 
Drawing a random sample $\bGamma^i \sim p(\bGamma)$, $(i=1:M),$ from the set of independent T priors yields the naïve Monte Carlo estimate  
\begin{equation*} 
\hat p(\y) = M^{-1} \sum_{i=1}^M |\I-\bGamma^i|_+ \ f(\y|\bGamma^i)
\end{equation*}
where   $f(\y|\bGamma)$ is the product of the univariate p.d.f.s for each series. 
A drawback of sampling from the prior is that the $\bGamma$ draws do not take into account the observed $\y$, so may be values that are in fact unlikely under the posterior.

\parablu{Sampling from the posterior $p(\bGamma|\y)$}. In order to produce more-probable draws of the $\bgamma_j$,   reorganization of the integrand in  
\eqn{app:pygamma} gives 
\begin{equation*} 
p(\y) =  f(\y) \int |\I-\bGamma|_+ \ \prod_{j=1}^q p(\bgamma_j|y_j,\y_{sp(j)})d\bGamma
\end{equation*}
where the terms in the product in the integrand are the p.d.f.s of the marginal {\em posterior} T distributions for parental coefficient subvectors in the set of series, and 
$f(\y) = \prod_{j=1}^q p_j(y_j)$ is the product of marginal predictive densities. 

A posterior random sample $\bGamma^i \sim p(\bGamma|\y)$, $(r=1:R)$,  then defines the estimate
\begin{equation}
\bar{p}(\y) = R^{-1} f(\y) \sum_{r=1}^R |\I-\bGamma^r|_+.
\end{equation}
Relative to the  \lq\lq naïve'' approach based on direct sampling from the prior, sampling from the posterior leads to Monte Carlo evaluations focused more heavily in the region supported by the data. This typically reduces the variability of the resulting Monte Carlo estimate of $p(\y)$.

%


\section{Sampling for Filtering in CFMs\label{app:smcdetails}}

\subsection{Main Details} 

As discussed in Section~\ref{sec:CFManalysis},   the importance sampling strategy for time $t$ prior-to-posterior updating in the CFM simply extends the existing SGDLM analysis to include the missing $\y_{e_0t}$ vector. The latter is sampled from a mixture of normals arising from conditional SGDLM theory.  
Important technical details for efficient computation using this strategy are noted here. For clarity, we drop the time subscript $t$ and conditioning information set $\cD_{t-1}$ throughout this section. 

Given $\bTheta,\bLambda$ the SGDLM distribution of $(\y_c',\y_e')'$ is $\tN(\balpha,\bSigma)$ where $\balpha=\A\bmu$ with $\A=(\I-\bGamma)^{-1},$ 
 and $\bSigma=\bOmega^{-1}$ where  $\bOmega=(\I-\bGamma')\bLambda(\I-\bGamma)$.   Calculation of the vector $\balpha$ requires a matrix inversion, and that of $\bOmega$ simple matrix products (and recognizing that $\bLambda$ is diagonal). We can avoid some 
 other expensive matrix calculations (in particular, we can avoid inverting $\bOmega$ directly to find $\bSigma$) by exploiting the known structure of $\bOmega.$   In conformably partitioned forms to define notation, we know that
$$ 
\bGamma = \begin{pmatrix}\bGamma_c&\bGamma_{ce}\\ \bGamma_{ec}&\bGamma_e\end{pmatrix}, \quad
\balpha = \begin{pmatrix}\balpha_c\\ \balpha_e\end{pmatrix},$$
$$\bSigma= \begin{pmatrix}\bSigma_c&\bSigma_{ce}\\ \bSigma_{ce}'&\bSigma_e\end{pmatrix} \quad\textrm{and}\quad 
\bOmega= \begin{pmatrix}\bOmega_c&\bOmega_{ce}\\ \bOmega_{ce}'&\bOmega_e\end{pmatrix}.
$$
Using standard theory, we have the following. 
\begin{itemize}
\item The margin $p(\y_c|\bTheta,\bLambda)$ is $\tN(\balpha_c,\bSigma_c)$ where $\bSigma_c^{-1} = \bOmega_c-\bOmega_{ce}\bOmega_e^{-1}\bOmega_{ce}'$. 
Define and compute $\B$ as the inverse of the lower triangular Cholesky factor of $\bOmega_e$, so that $\bOmega_e^{-1} = \B'\B$.  Further, define and compute $\F=\bOmega_{ce}\B'$.   It follows that $\bSigma_c^{-1} = \bOmega_c-\F\F'$.  This is easily computed, as is its determinant, to provide the terms for direct computation of  the log p.d.f. $\ell = \log\{p(\y_c|\bTheta,\bLambda)\}$ (up to an irrelevant constant)  via
$\ell =\log(|\bSigma_c^{-1}|)/2 -  (\y_c-\balpha_c)' \bSigma_c^{-1}(\y_c-\balpha_c)/2.$     
\item The conditional $p(\y_e|\y_c,\bTheta,\bLambda)$ is normal with mean vector $\balpha_e-\bOmega_e^{-1}\bOmega_{ce}'(\y_c-\balpha_c)$ and variance matrix
$\bOmega_e^{-1}.$  The MC strategy samples this distribution, and evaluating its parameters efficiently impacts on overall computational efficiency. With $\B,\F$ as defined and already computed above, it easily follows that a single MC sample from $p(\y_e|\y_c,\bTheta,\bLambda)$ is efficiently given by
$\y_e = \balpha_e - \B'\{ \F'(\y_c-\balpha_c) + \z\}$ where $\z\sim\tN(\bzero,\I).$ 
\end{itemize} 
Now,  the normal mixture to be sampled has the form 
 $$ p(\y_{e_0}|\y_{c}) = \sum_{r=1}^R  \pi_{r}(\y_{c}) \ 
         p(\y_{e_0}|\y_{c},\bTheta^r,\bLambda^r)
 $$ 
where each $\pi_{r}(\y_{c}) \propto p(\y_{c}|\bTheta^r,\bLambda^r)$.  The above general theoretical results apply to each component $r$ of this mixture. First, these define efficient computation of the mixture weights $\pi_{r}(\y_{c})$.  Sampling then generates a full set of mixture indices via a multinomial draw on cells $1:R$ with probabilities $\pi_{\seq 1R}(\y_{c}).$  Given each sampled component, the conditional normal results above apply to sample $\y_{e_0}.$ Note that 
these matrix calculations will be made only once on each component $r$ drawn (however many times) in the multinomial sample. 
 
\subsection{Additional Theoretical Aspects}
This strategy assumes that the MC mixture form of the p.d.f. for $\y_{e_0t}|\y_{ct},\cD_{t-1}$ 
in \eqn{MCmixpy0givenyc} is in fact the theoretically exact conditional.  As noted, a large time $t-1$ MC sample size $R$ is important in underpinning this. 

To clarify this more fundamentally, relabel the mixture of \eqn{MCmixpy0givenyc} as 
$g(\y_{e_0t}|\y_{ct},\cD_{t-1}),$ explicitly recognizing it as an approximation to the exact 
conditional $p(\y_{e_0t}|\y_{ct},\cD_{t-1}).$  The latter is, of course, not available in any useful analytic form, or for direct simulation.   The exact joint posterior for the missing data and parameters is easily seen to have the form
\begin{equation} p(\y_{e_0t},\bTheta_t,\bLambda_t|\y_{ct},\cD_{t-1}) \propto 
             p(\y_{e_0t}|\y_{ct},\cD_{t-1}) \ |\I-\bGamma_t|_+ \ \tip_t(\bTheta_t,\bLambda_t|\y_{ct},\y_{e_0t},\cD_{t-1}).
\end{equation}
Use of the MC mixture in the importance sampling proposal implies the joint proposal p.d.f. of the form 
\begin{equation}g(\y_{e_0t},\bTheta_t,\bLambda_t|\y_{ct},\cD_{t-1}) \propto 
             g(\y_{e_0t}|\y_{ct},\cD_{t-1}) \tip_t(\bTheta_t,\bLambda_t|\y_{ct},\y_{e_0t},\cD_{t-1}).
\end{equation}
The implied importance weight is then proportional to 
$$ |\I-\bGamma_t|_+ \ \frac{p(\y_{e_0t}|\y_{ct},\cD_{t-1})}  
                       {g(\y_{e_0t}|\y_{ct},\cD_{t-1})}.$$  
This makes transparent the fact that assuming the MC mixture form for the marginal posterior for the missing data leads to the usual SGDLM IS weight based on the determinantal term. It also clearly shows: (i) how intractable a direct IS approach is, since the target marginal $p(\y_{e_0t}|\y_{ct},\cD_{t-1})$ cannot be evaluated; and 
(ii) that even if it could be, the computational burden would escalate due to the need to evaluate the mixture p.d.f. in the denominator of the weight.

\section{Factor Identification\label{app:factorID}} 
The mapping of posterior inferences on the SGDLM parameters to the implied factor components-- exemplified in Section~\ref{sec:GDPsparsityandfactors}--  requires attention to 
inherent ambiguities in factor signs, ordering and matching between time points. These are 
resolved as follows.  The SVD is applied to  each of the posterior Monte Carlo samples of $\bGamma_t$ at each time $t$. \lq\lq Permuting factors"  then refers to applying a chosen permutation to the rows of $\fbF_t$, the diagonal of $\D_t$ and the columns of $\fbA_t$ generated from the SVD.  The first step is to permute factors if needed to match the zero/non-zero pattern of $\S_t$ with a chosen reference pattern; the latter is arbitrary, here chosen simply for clarity and aesthetics of resulting figures. The second step
 multiplies each row of $\fbF_t$ and the corresponding column of $\fbA_t$ by $\pm 1$ so that the largest element in absolute value 
 in the row is positive. The third step applies (only) to each parental set that underlies more than one factor. 
 Within each such $\cP_h$  this permutes the 
 factors in order so that the first has the largest score in the relevant row of $\S_t$ 
 from among this subset of  series in $\cP_h,$  the second has the largest ignoring the first, and so forth. 
 These second and third steps relate to 
 ordering of factors \lq\lq founder'' variables in~\cite{Carvalho2008},  extended and referred to as \lq\lq pivot'' variables in~\cite{FruhwirthSetal2024}.

 \end{appendices}

\if0\blind
	{\section*{Acknowledgements}
The research reported here developed while Luke Vrotsos was a PhD student in Statistical Science at Duke University ({\small 2022-2026}). 
	} \fi  
%

%

\small
\bibliography{VrotsosWestSGDLMs2026}

@Manual{SAS-SGDLMs,
  title  = {The {DYNAMICLINEAR} Procedure.},
  author = {C. Xu}, 
  year   = {2024},
  organization = {SAS},
  url    ={https://documentation.sas.com/doc/en/pgmsascdc/v_068/casecon/casecon_dynamiclinear_overview.htm},
  note ={Econometrics Software: https://documentation.sas.com/},
}

@Article{West2020Akaike,
  author  = {M. West},
  journal = {Annals of the Institute of Statistical Mathematics},
  title   = {Bayesian forecasting of multivariate time series: {S}calability, structure uncertainty and decisions (with discussion)},
  year    = {2020},
  pages   = {1--44},
  volume  = {72}, 
  doi     = {10.1007/s10463-019-00741-3},
}

@Article{BolfarineEtAl2024,
  author   = {Henrique Bolfarine and Carlos M. Carvalho and Hedibert F. Lopes and Jared S. Murray},
  journal  = {Bayesian Analysis},
  title    = {Decoupling shrinkage and selection in {G}aussian linear factor analysis},
  year     = {2024},
  pages    = {181--203},
  volume   = {19},
  doi      = {10.1214/22-BA1349},
}

@Article{BeyelerKaufmann2021,
  author  = {Beyeler, Simon and Kaufmann, Sylvia},
  year = {2021},
  title = {Reduced-form factor augmented {VAR}- {E}xploiting sparsity to include meaningful factors},
  journal = {Journal of Applied Econometrics},
   volume ={36},
   pages={989--1012},
   }

@Article{abadie2015comparative,
  author    = {Abadie, Alberto and Diamond, Alexis and Hainmueller, Jens},
  journal   = {American Journal of Political Science},
  title     = {Comparative politics and the synthetic control method},
  year      = {2015},
  pages     = {495--510},
  volume    = {59},
  publisher = {Wiley Online Library},
  doi = {10.1111/ajps.12116},
}

@Article{KevinLiEtAlCausalMVTS2024,
  author  = {K. Li and G. Tierney and C. Hellmayr and M. West},
  journal = {Applied Stochastic Models in Business and Industry},
  title   = {Compositional dynamic modelling for counterfactual prediction in multivariate time series},
  year    = {2025},
  pages   = {e2908},
  volume  = {41},
  doi     = {10.1002/asmb.2908},
  url     = {https://doi.org/10.1002/asmb.2908},
}

@Book{WestHarrison1997,
  author       = {M. West and P. J. Harrison},
  publisher    = {Springer},
  title        = {Bayesian Forecasting and Dynamic Models},
  year         = {1997},
  edition      = {2nd},
  creationdate = {2010-03-16T00:00:00},
  key          = {WestHarrison.YellowBook.1997},
  url          = {http://www.stat.duke.edu/~mw/West&HarrisonBook/},
  doi ={10.1007/b98971},
}

@Article{West1986,
  author  = {M. West and P. J. Harrison},
  journal = {Journal of the American Statistical Association},
  title   = {Monitoring and adaptation in {B}ayesian forecasting models},
  year    = {1986},
  pages   = {741--750},
  volume  = {81},
  doi = {10.1080/01621459.1986.10478331},
}

@Article{TierneyEtAl2024,
  author  = {G. Tierney and C. Hellmayr and K. Li and G. Barkimer and M. West},
  journal = {Bayesian Analysis},
  title   = {Multivariate {B}ayesian dynamic modeling for causal prediction},
  year    = {2024},
  doi     = {10.1214/24-BA1493},
  url     = {https://doi.org/10.1214/24-BA1493},
}

@Article{TierneyEtAl2024Supplement,
  author  = {G. Tierney and C. Hellmayr and K. Li and G. Barkimer and M. West},
  journal = {Bayesian Analysis (supplementary material)},
  title   = {Multivariate {B}ayesian dynamic modeling for causal prediction: {M}ore on models, data and analyses},
  year    = {2024},
  doi     = {10.1214/24-BA1493SUPP},
  url     = {https://doi.org/10.1214/24-BA1493SUPP},
}

@Article{GruberWest2016,
  author  = {L. F. Gruber and M. West},
  journal = {Bayesian Analysis},
  title   = {{GPU}-accelerated {B}ayesian learning and forecasting in simultaneous graphical dynamic linear models},
  year    = {2016},
  pages   = {125--149},
  volume  = {11},
  doi     = {10.1214/15-BA946},
  url     = {http://projecteuclid.org/euclid.ba/1425304898},
}

@Article{GruberWest2017,
  author       = {L. F. Gruber and M. West},
  journal      = {Econometrics and Statistics},
  title        = {Bayesian forecasting and scalable multivariate volatility analysis using simultaneous graphical dynamic linear models},
  year         = {2017},
  pages        = {3--22},
  volume       = {3},
  comment      = {Published online March 12, 2017},
  creationdate = {2017-09-22T00:00:00},
  doi          = {10.1016/j.ecosta.2017.03.003},
  url          = {http://www.sciencedirect.com/science/article/pii/S2452306217300163},
}

@Article{Kyakutwika2023,
  author  = {Nelson Kyakutwika and Bruce Bartlett},
  journal = {Investment Analysts Journal},
  title   = {Bayesian forecasting of stock returns on the {JSE} using simultaneous graphical dynamic linear models},
  year    = {2024},
  pages   = {1--20},
  doi     = {10.1080/10293523.2024.2312712},
}

@Article{GriveauEtAl2020,
  author  = {Griveau-Billion, Th{\'{e}}ophile and Calderhead, B.},
  journal = {Quantitative Finance},
  title   = {Efficient computation of mean reverting portfolios using cyclical coordinate descent},
  year    = {2021},
  pages   = {673--684},
  volume  = {21},
  doi     = {10.1080/14697688.2020.1803497},
}

@Article{TiaoTsay1989,
  author  = {G. C. Tiao and R. S. Tsay},
  journal = {Journal of the Royal Statistical Society: Series B (Statistical Methodology)},
  title   = {Model specification in multivariate time series},
  year    = {1989},
  pages   = {157--195},
  volume  = {51},
  url     = {https://rss.onlinelibrary.wiley.com/doi/10.1111/j.2517-6161.1989.tb01756.x},
  doi = {10.1111/j.2517-6161.1989.tb01756.x},
}

@Article{Yoshida2010,
  author  = {R. Yoshida and M. West},
  journal = {Journal of Machine Learning Research},
  title   = {Bayesian learning in sparse graphical factor models via annealed entropy},
  year    = {2010},
  pages   = {1771--1798},
  volume  = {11},
  note ={http://jmlr.org/papers/v11/yoshida10a.html},
}

@Article{LopesCarvalho07,
  author  = {Lopes, H. F. and C. M. Carvalho},
  journal = {Journal of Statistical Planning and Inference},
  title   = {Factor stochastic volatility with time varying loadings and {M}arkov switching regimes},
  year    = {2007},
  pages   = {3082--3091},
  volume  = {137},
  doi ={10.1016/j.jspi.2006.06.047},
}

@Article{Lopesetal2022,
  author  = {Hedibert F. Lopes and Robert E. McCulloch and Ruey S. Tsay},
  journal = {Journal of Econometrics},
  title   = {Parsimony inducing priors for large scale state–space models},
  year    = {2022},
  pages   = {39--61},
  volume  = {230},
  doi     = {10.1016/j.jeconom.2021.11.005},
}

@Article{Primiceri05,
  author  = {Primiceri, G. E.},
  journal = {Review of Economic Studies},
  title   = {Time varying structural vector autoregressions and monetary policy},
  year    = {2005},
  pages   = {821--852},
  volume  = {72},
  doi = {https://doi.org/10.1111/j.1467-937X.2005.00353.x},
}

@Article{FruhwirthSetal2024,
  author  = {Sylvia Fr{\"u}hwirth-Schnatter and Darjus Hosszejni and Hedibert Freitas Lopes},
  journal = {Bayesian Analysis},
  title   = {Sparse {B}ayesian factor analysis when the number of factors is unknown},
  year    = {2024}, 
  doi     = {10.1214/24-BA1423},
}

@Article{Kaufmann2019,
  author  = {Sylvia Kaufmann and Christian Schumacher},
  journal = {Journal of Econometrics},
  title   = {Bayesian estimation of sparse dynamic factor models with order-independent and ex-post mode identification},
  year    = {2019},
  pages   = {116--134},
  volume  = {210},
  doi     = {10.1016/j.jeconom.2018.11.008},
}

@Article{KoopKorobilis2013,
  author  = {Gary Koop and Dimitris Korobilis},
  journal = {Journal of Econometrics},
  title   = {Large time-varying parameter {VARs}},
  year    = {2013},
  pages   = {185--198},
  volume  = {177},
  doi     = {https://doi.org/10.1016/j.jeconom.2013.04.007},
  url     = {https://www.sciencedirect.com/science/article/pii/S0304407613000845},
}

@Article{KoopKorobilis2010,
  author  = {Gary Koop and Dimitris Korobilis},
  journal = {Foundations and Trends in Econometrics},
  title   = {Bayesian multivariate time series methods for empirical macroeconomics},
  year    = {2010},
  issn    = {1551-3076},
  pages   = {267--358},
  volume  = {3},
  doi     = {10.1561/0800000013},
}

@Article{BanburaEtAl2010,
  author  = {Marta Ba{\'{n}}bura and Domenico Giannone and Lucrezia Reichlin},
  journal = {Journal of Applied Econometrics},
  title   = {Large {B}ayesian vector autoregressions},
  year    = {2010},
  pages   = {71--92},
  volume  = {25},
  doi = {10.1002/jae.1137},
}

@Article{NakajimaWest2013JFE,
  author  = {J. Nakajima and M. West},
  journal = {Journal of Financial Econometrics},
  title   = {Bayesian dynamic factor models: {L}atent threshold approach},
  year    = {2013},
  pages   = {116--153},
  volume  = {11},
  doi     = {10.1093/jjfinec/nbs013},
  url     = {http://www.stat.duke.edu/~mw/MWextrapubs/NakajimaWest2013JFE.pdf},
}

@Article{NakajimaWest2017BJPS,
  author  = {J. Nakajima and M. West},
  journal = {Brazilian Journal of Probability and Statistics},
  title   = {Dynamics and sparsity in latent threshold factor models: {A} study in multivariate {EEG} signal processing},
  year    = {2017},
  pages   = {701--731},
  volume  = {31},
  doi     = {10.1214/17-BJPS364},
  url     = {https://projecteuclid.org/euclid.bjps/1513328764},
}

@Article{NakajimaWest2013JBES,
  author  = {J. Nakajima and M. West},
  journal = {Journal of Business and Economic Statistics},
  title   = {Bayesian analysis of latent threshold dynamic models},
  year    = {2013},
  pages   = {151--164},
  volume  = {31},
  doi     = {10.1080/07350015.2012.747847},
  url     = {http://www.stat.duke.edu/~mw/MWextrapubs/NakajimaWest2013JBES.pdf},
}

@Article{NakajimaWest2015DSP,
  author       = {J. Nakajima and M. West},
  journal      = {Digital Signal Processing},
  title        = {Dynamic network signal processing using latent threshold models},
  year         = {2015},
  pages        = {6--15},
  volume       = {47},
  creationdate = {2016-04-11T00:00:00},
  doi          = {10.1016/j.dsp.2015.04.008},
  url          = {http://www.stat.duke.edu/~mw/MWextrapubs/NakajimaWest2015DSP.pdf},
}

@Article{Carvalho2008,
  author  = {C. M. Carvalho and J. E. Lucas and Q. Wang and J. Chang and J. R. Nevins and M. West},
  journal = {Journal of the American Statistical Association},
  title   = {High-dimensional sparse factor modelling-- {A}pplications in gene expression genomics},
  year    = {2008},
  pages   = {1438--1456},
  volume  = {103},
}

@Article{klosner_comparative_2018,
  author  = {Klößner, Stefan and Kaul, Ashok and Pfeifer, Gregor and Schieler, Manuel},
  journal = {Swiss Journal of Economics and Statistics},
  title   = {Comparative politics and the synthetic control method revisited: {A} note on {Abadie} et al. (2015)},
  year    = {2018},
  pages   = {11},
  volume  = {154},
  doi     = {10.1186/s41937-017-0004-9},
}

@Article{ZhouNakajimaWest2014IJF,
  author       = {X. Zhou and J. Nakajima and M. West},
  journal      = {International Journal of Forecasting},
  title        = {Bayesian forecasting and portfolio decisions using dynamic dependent sparse factor models},
  year         = {2014},
  pages        = {963--980},
  volume       = {30},
  creationdate = {2016-04-11T00:00:00},
  doi          = {http://dx.doi.org/10.1016/j.ijforecast.2014.03.017},
  url          = {http://www.stat.duke.edu/~mw/MWextrapubs/ZhouNakajimaWest2014.pdf},
}

@InCollection{Simovici2014,
  author    = {Simovici, Dan A. and Djeraba, Chabane},
  booktitle = {Mathematical Tools for Data Mining: Set Theory, Partial Orders, Combinatorics},
  publisher = {Springer},
  title     = {Spectral properties of matrices},
  year      = {2014},
  pages     = {347--397},
  doi       = {10.1007/978-1-4471-6407-4\_7},
  url       = {https://doi.org/10.1007/978-1-4471-6407-4_7},
}

@Book{PradoFerreiraWest2021,
  author    = {R. Prado and M. A. R. Ferreira and M. West},
  publisher = {Chapman \& Hall/CRC Press},
  title     = {Time Series: Modeling, Computation \& Inference},
  year      = {2021},
  edition   = {2nd},
  isbn = {978-1498747028},
  doi = {10.1201/9781351259422},
}

@Book{HornJohnson2012,
  author    = {R. Horn and C. Johnson},
  publisher = {Cambridge University Press},
  title     = {Matrix Analysis},
  year      = {2012},
  edition   = {2nd},
  doi ={10.1017/CBO9780511810817},
}

@Article{marimont_system_1969,
  author   = {Marimont, Rosalind B.},
  journal  = {The Bulletin of Mathematical Biophysics},
  title    = {System connectivity and matrix properties},
  year     = {1969},
  issn     = {1522-9602},
  pages    = {255--274},
  volume   = {31},
  doi      = {10.1007/BF02477005},
  language = {en},
  url      = {https://doi.org/10.1007/BF02477005},
  urldate  = {2024-02-26},
}

@Book{cvetkovic_spectra_1980,
  author     = {Cvetkovic, Dragos and Doob, Michael and Sachs, Horst},
  publisher  = {Academic Press},
  title      = {Spectra of {Graphs}: {Theory} and {Application}},
  year       = {1980},
  address    = {New York}, 
  note ={ISBN 9780121951504 / 0121951502},   
}

@Article{fang_low-rank_2024,
  author  = {Fang, Zhuangyan and Zhu, Shengyu and Zhang, Jiji and Liu, Yue and Chen, Zhitang and He, Yangbo},
  journal = {IEEE Transactions on Neural Networks and Learning Systems},
  title   = {On low-rank directed acyclic graphs and causal structure learning},
  year    = {2024},
  pages   = {4924--4937},
  volume  = {35},
  doi     = {10.1109/TNNLS.2023.3273353},
}

@Article{pang_bayesian_2022,
  author  = {Pang, Xun and Liu, Licheng and Xu, Yiqing},
  journal = {Political Analysis},
  title   = {A {B}ayesian alternative to synthetic control for comparative case studies},
  year    = {2022},
  issn    = {1047-1987, 1476-4989},
  pages   = {269--288},
  volume  = {30},
  doi = {10.1017/pan.2021.22},
}

@Article{Brodersen2015,
  author   = {Brodersen, Kay H. and Gallusser, Fabian and Koehler, Jim and Remy, Nicolas and Scott, Steven L.},
  journal  = {The Annals of Applied Statistics},
  title    = {Inferring causal impact using {Bayesian} structural time-series models},
  year     = {2015},
  issn     = {1932-6157},
  pages    = {247--274},
  volume   = {9},
  doi      = {10.1214/14-AOAS788},
  language = {en},
  urldate  = {2024-05-07},
}

@Article{Abadie2003,
  author  = {Abadie, Alberto and Gardeazabal, Javier},
  journal = {American Economic Review},
  title   = {The economic costs of conflict: {A} case study of the {Basque} {Country}},
  year    = {2003},
  pages   = {113--132},
  volume  = {93},
  doi = {10.1257/000282803321455188},
}

@Article{MenchettiBojinov2022,
  author  = {Menchetti, Fiammetta and Bojinov, Iavor},
  journal = {The Annals of Applied Statistics},
  title   = {Estimating the effectiveness of permanent price reductions for competing products using multivariate {Bayesian} structural time series models},
  year    = {2022},
  pages   = {414--435},
  volume  = {16},
  doi     = {10.1214/21-AOAS1498},
}

@Article{AntonelliBeck2023,
  author  = {Antonelli, Joseph and Beck, Brenden},
  journal = {Journal of the Royal Statistical Society: Series A (Statistics in Society)},
  title   = {Heterogeneous causal effects of neighbourhood policing in {New} {York} {City} with staggered adoption of the policy},
  year    = {2023},
  issn    = {0964-1998},
  pages   = {772--787},
  volume  = {186},
  doi     = {10.1093/jrsssa/qnad058},
}

@Manual{pySGDLM,
  title       = {SGDLM in python.},
  author      = {N. Kyakutwika},
  note        = {Software: https://github.com/nelsonkyakutwika/SGDLM}, 
  year        = {2023},
  institution = {Stellenbosch University, South Africa},
}

@Manual{GruberSGDLMrCode,
  title  = {{GPU}-accelerated software for online learning of the simultaneous graphical {DLM}.},
  author = {L. F. Gruber},
  note   = {Software: https://github.com/lutzgruber/gpuSGDLM}, 
  year   = {2018},
}

@Article{Jones2005a,
  author       = {B. Jones and A. Dobra and C. M. Carvalho and C. Hans and C. Carter and M. West},
  journal      = {Statistical Science},
  title        = {Experiments in stochastic computation for high-dimensional graphical models},
  year         = {2005},
  pages        = {388--400},
  volume       = {20},
  creationdate = {2010-03-16T00:00:00},
  key          = {Jones.StatSci.2005},
  url          = {https://projecteuclid.org/euclid.ss/1137076659},
  doi = {10.1214/088342305000000304}, 
}

@Article{Jones2005,
  author       = {B. Jones and M. West},
  journal      = {Biometrika},
  title        = {Covariance decomposition in undirected {G}aussian graphical models},
  year         = {2005},
  pages        = {779--786},
  volume       = {92},
  creationdate = {2010-03-16T00:00:00},
  doi          = {10.1093/biomet/92.4.779},
  key          = {Jones.Biometrika.2005},
  url          = {http://www.stat.duke.edu/~mw/MWextrapubs/Jones2005.pdf},
}

@Article{LavineLindonWest2019avs,
  author  = {I. Lavine and M. Lindon and M. West},
  journal = {Bayesian Analysis},
  title   = {Adaptive variable selection for sequential prediction in multivariate dynamic models},
  year    = {2021},
  pages   = {1059--1083},
  volume  = {16},
  doi     = {10.1214/20-BA1245},
  url     = {https://doi.org/10.1214/20-BA1245},
}

@Article{Hans2007a,
  author       = {C. Hans and A. Dobra and M. West},
  journal      = {Journal of the American Statistical Association},
  title        = {Shotgun stochastic search in regression with many predictors},
  year         = {2007},
  pages        = {507--516},
  volume       = {102},
  creationdate = {2010-03-16T00:00:00},
  key          = {Hans.SSS.ISBA.2007},
  url          = {http://www.stat.duke.edu/~mw/MWextrapubs/Hans2007a.pdf},
  doi = {10.1198/016214507000000121},
}

@Article{TallmanWest2023,
  author  = {E. Tallman and M. West},
  journal = {Journal of the Royal Statistical Society (Ser. B)},
  title   = {Bayesian predictive decision synthesis},
  year    = {2023},
  pages   = {340--363},
  volume  = {86},
  doi     = {10.1093/jrsssb/qkad109},
  url     = {https://doi.org/10.1093/jrsssb/qkad109},
}

@InCollection{TallmanWest2024,
  author    = {E. Tallman and M. West},
  booktitle = {Recent Developments in Bayesian Econometrics and Their Applications},
  publisher = {Springer},
  title     = {Predictive decision synthesis for portfolios: {B}etting on better models},
  year      = {2025},
  pages = {223--249},
  editor    = {S. Mazur and P. {\"O}sterhol},
  url = {https://doi.org/10.1007/978-3-032-00110-8_10}, 
  doi={10.1007/978-3-032-00110-8\_10},
  isbn = {978-3-032-00110-8}
}

@Article{LavineCronWest2020factorDGLMs,
  author  = {I. Lavine and A. J. Cron and M. West},
  journal = {Journal of Computational and Graphical Statistics},
  title   = {Bayesian computation in dynamic latent factor models},
  year    = {2022},
  pages   = {651--665},
  volume  = {31},
  doi     = {10.1080/10618600.2021.2021208},
  url     = {https://www.tandfonline.com/doi/abs/10.1080/10618600.2021.2021208?src=&journalCode=ucgs20},
}

@Article{ZhaoXieWest2016ASMBI,
  author  = {Z. Y. Zhao and M. Xie and M. West},
  journal = {Applied Stochastic Models in Business and Industry},
  title   = {Dynamic dependence networks: {F}inancial time series forecasting and portfolio decisions (with discussion)},
  year    = {2016},
  pages   = {311--339},
  volume  = {32},
  comment = {Reply to invited discussion: pages 336--339. doi: 10.1002/asmb.2169},
  doi     = {10.1002/asmb.2161},
  url     = {https://onlinelibrary.wiley.com/doi/abs/10.1002/asmb.2161},
}
\bibliographystyle{chicago} 
\normalsize

\end{document}